\documentclass{PoS}

\title{T-Duality and Doubling of the Isotropic Rigid Rotator}

\ShortTitle{T-Duality and Rotator}

\author{Francesco Bascone\\
       Dipartimento di Fisica ``E. Pancini'', Universit\`a di Napoli Federico II and  INFN-Sezione di Napoli, Complesso Universitario  di Monte S. Angelo Edificio 6, via Cintia, 80126 Napoli, Italy.\\
       E-mail: \email{francesco.bascone@na.infn.it}}

\author{Vincenzo E. Marotta \\
     Department of Mathematics, Heriot-Watt University Colin Maclaurin Building, Riccarton, Edinburgh EH14 4AS, U.K. \\
E-mail: \email{vm34@hw.ac.uk}}
\author{ \speaker{Franco Pezzella} %\thanks{A footnote may follow.}
\\
         INFN - sezione di Napoli, Complesso Universitario di Monte S. Angelo ed. 6, via Cintia - 80126 Napoli, Italy\\
        E-mail: \email{pezzella@na.infn.it}}
        
   \author{Patrizia Vitale\\
       Dipartimento di Fisica ``E. Pancini'', Universit\`a di Napoli Federico II and  INFN-Sezione di Napoli, Complesso Universitario  di Monte S. Angelo Edificio 6, via Cintia, 80126 Napoli, Italy.\\
       E-mail: \email{patrizia.vitale@na.infn.it}}

\abstract{After reviewing some of the fundamental aspects of Drinfel'd doubles and Poisson-Lie T-duality, we describe the three-dimensional isotropic rigid rotator on $SL(2,\mathbb{C})$ starting from a non-Abelian deformation of the natural carrier space of its Hamiltonian description on $T^*SU(2) \simeq SU(2) \ltimes \mathbb{R}^3$.  A new model is then introduced on the dual group $SB(2,\mathbb{C})$, within the Drinfel'd double description of $SL(2,\mathbb{C})=SU(2) \bowtie SB(2,\mathbb{C})$.  The two models are analyzed from the Poisson-Lie duality point of view,  and a doubled generalized action is built with $TSL(2,\mathbb{C})$ as carrier space. The aim is to explore within a simple case the relations between Poisson-Lie symmetry, Doubled Geometry and Generalized Geometry. In fact,  all the mentioned structures are discussed, such as a Poisson realization of the $C$-brackets for the generalized bundle $T \oplus T^*$ over $SU(2)$ from the Poisson algebra of the generalized model. 
The two dual models exhibit many features  of Poisson-Lie duals and from the generalized action both of them can be respectively recovered by gauging one of its symmetries.  %The same construction is  then sketched for the Principal Chiral Model. %However, the latter has the purpose of making T-duality manifest in the target space low-energy effective theory. DFT can be considered indeed a manifestly T-dual invariant reformulation of Supergravity. Many attempts have been done in the direction of generalizing the DFT description to Poisson-Lie T-duality, which is a generalization of the standard Abelian and non-Abelian concepts of T-dualities, see e.g. \cite{Hassler2017}. 

        }

\FullConference{Corfu Summer Institute 2018 "School and Workshops on Elementary Particle Physics and Gravity"\\
		(CORFU2018)\\
		31 August - 28 September, 2018\\
		Corfu, Greece}

\begin{document}

%\section{...}
\section{Introduction}
\label{secintro}

T-duality is a peculiar feature of strings, since it is inherently linked to their extended nature and, therefore, cannot subsist for point particles. It originally emerges out of compactifying the target space along some directions.  Indeed, the one-dimensionality of a string allows it to wrap around non-contractible cycles, resulting in {\em winding mode} contributions to its dynamics in addition to the ones of momentum modes  familiar from particle dynamics, quantized in integer units. T-duality relates these two kinds of modes. It implies that, in many cases, two different geometries for the extra dimensions are physically equivalent, i.e. string physics at a very small scale cannot be distinguished from the one at a large scale, as it is shown in the simplest case of compactification of a spatial coordinate on a circle of radius $R$. Here, T-duality is encoded by the simultaneous transformations of $ R \leftrightarrow \frac{\alpha'}{R}$  and $p_a \leftrightarrow  w^a$, where $p_a$ is the quantized momentum of the string and $w^a$ is its winding mode. Under such transformations, the string coordinate  along a compact dimension $X^a$, sum of the left and right movers, is transformed into the T-dual coordinate $\tilde{X}_{a}$, defined by their difference. The winding mode $w^{a}$ plays with respect to $\tilde{X}_a$ the same role as $p_a$  does with respect to the coordinate $X^{a}$. This leads to a duality between two different target spaces  with the string theories  defined on them resulting to be equivalent, as can be seen from the mass spectrum analysis. %For example from the target space correspondence we understand that strings physics at very small scale cannot be distinguished from one at a large scale. 
%This is also important because it allows to perform calculations of quantities in the dual theory where the quantities may be easier to compute. 
 More generally, T-duality allows to build new string backgrounds which could not be addressed otherwise and generally go under the name of  \textit{non-geometric backgrounds} (see e.g. \cite{Plauschinn2018} for a recent review on the subject). Furthermore, together with S-duality and U-duality, it lies at the heart of relating the five different superstring theories that turn out to be seen as low-energy limits  of the so-called M-theory.

On a $d$-dimensional torus T-duality is described by an $O(d,d;\mathbb{Z})$ transformation that  is a symmetry of the Hamiltonian but not of the action of the theory.  
This is actually reminiscent of a more general symmetry. 

Indeed, one can observe that,   already at the classical level, the symmetry under the indefinite orthogonal group $O(D,D; \mathbb{R})$  - in the following we will refer to it as $O(D,D)$ - naturally appears in the Hamiltonian description of the bosonic string in a $D$-dimensional pseudo-Riemannian target space $M$ with constant background $(G,B)$, being $G$ the target space metric and $B$ the Kalb-Ramond field. Such string is described by the well-known two-dimensional non-linear sigma model  action:
\vspace{.05cm}
\begin{eqnarray}
S[X; G, B]  =  -\frac{1}{4 \pi \alpha'} \int_{\Sigma} d\tau \, d\sigma \left[ h^{\alpha \beta} \partial_{\alpha} X^{a} \partial_{\beta} X^{b} G_{ab} + \epsilon^{\alpha \beta} \partial_{\alpha} X^{a} \partial_{\beta} X^{b}  B_{ab} \right]  \equiv   \int_{\Sigma} d\tau \, d\sigma L \label{action}
\end{eqnarray}
\vspace{.05cm}
where the two-dimensional world-sheet  $\Sigma$ $(\alpha, \beta=0,1)$, with metric $h= \mbox{diag}(-1,1)$ in the conformal gauge, is embedded into $M$ ($a,b=0, \dots, D-1$).  The convention $\epsilon^{01} = - \epsilon^{10} = -1$ is adopted from now on. 

The Hamiltonian density $H$ can be derived from the Lagrangian density $L$ in eq. (\ref{action}) by a Legendre transformation with respect to $\partial_{0} X^{a}$.  The canonical momentum $P_{a}$ is given by:
\begin{equation}
P_{a} \equiv \frac{\partial L}{\partial \partial_{0} X^{a}} = \frac{1}{ 2 \pi \alpha'} \left( G_{ab} \partial_{0} X^{b} + B_{ab} \partial_{1} X^{b} \right) \,\,\,  \nonumber 
\end{equation}
with inverse
\begin{equation}
\partial_0 X^a= 2\pi\alpha'({ G^{-1}})^{ab}P_b- ({G^{-1}})^{ac} B_{cb}\partial_1 X^b  \label{delX0}
\end{equation}
The result is:
\begin{eqnarray}
H & \equiv & P \cdot  \partial_{0} X  - L \\ \nonumber 
&= & \frac{1}{ 4 \pi \alpha'} \left( \partial_{0} X \cdot \partial_{0} X+ \partial_{1} X \cdot \partial_{1} X \right)|_{(\partial_{0} X)(P)} \,\,  \label{H1}
\end{eqnarray} 
which, by means of  eq. (\ref{delX0}), yields 
\begin{equation}
H=\frac{1}{4 \pi \alpha'} \left(
\begin{array}{c}
  \partial_{1}X  \\
  2 \pi \alpha' P
   \end{array} \right)^T \mathcal{H}(G,B) \left( \begin{array}{c}
  \partial_{1}X  \\
  2 \pi \alpha' P
   \end{array} \right)   \label{hamden2}
\end{equation}
and the so-called \textit{generalized metric} 
\begin{equation}
\mathcal{H}(G,B)= \left( \begin{array}{cc}
  G-BG^{-1}B & BG^{-1} \\
  -G^{-1}B & G^{-1} 
 \end{array}  \right)
\end{equation}
 has been introduced.

Hence, $H$ in eq. (\ref{hamden2})  is proportional to the squared length of the following $2D$-dimensional generalized vector  field $A_p \in TM \oplus T^*M$:
\begin{equation}
A_P (X) \equiv \partial_{1} X^{a} \partial_{a}+2 \pi \alpha' P_{a} dX^{a}
\end{equation}
as measured by the generalized metric, i.e. $H= \frac{1}{4 \pi \alpha'} \,  A_P^{T} \,\,  \mathcal{H} \,\, A_P$.

The Hamiltonian density $H$  in eq. (\ref{H1}) results to be invariant under %$\partial_{0} X \leftrightarrow \partial_{1} X$ or, more fundamentally, under $\tau \leftrightarrow \sigma$, is transferred to its invariance under 
the exchange $\partial_{1} X \leftrightarrow 2 \pi \alpha' P$ in eq. (\ref{hamden2}), realized by the transformation 
\begin{equation}
A_P \rightarrow \tilde A_P = \eta A_P    \label{etatra}
\end{equation}
involving the $O(D,D)$-invariant metric:
\begin{equation}
\eta= \left( \begin{array}{cc}
  0 & 1 \\
  1 & 0     \label{eta}
 \end{array} \right)
\end{equation}
if also the metric and the Kalb-Ramond field are transformed according to the transformation of $\mathcal{H}$ as:
\begin{equation}
\mathcal{H} \rightarrow \mathcal{H}' = {\cal H}^{-1} = \eta \,  {\cal H} \, \eta  \,\,\, .  \nonumber
\end{equation}
%\textcolor{red}{\bf Mi  pare infatti che le equazioni del moto siano invarianti rispetto allo scambio $ P \leftrightarrow \partial_1 X$, ma non l'Hamiltoniana. A meno di scambiare la metrica generalizzata con la sua inversa, come in realta` e` spiegato dopo. Se siete d'accordo l'affermazione va corretta con l'aggiunta della frase in blu di cui sopra.}
Actually, it is easy to check  % that $H$ is left invariant by the transformation (\ref{etatra}) made on the generalized vector $A_P$ if, simultaneously, ${\cal H} \leftrightarrow {\cal H}^{-1}$. 
that the invariance of $H$ under $A_P\rightarrow {\mathcal T} A_P$ holds for any element ${\cal T} \in O(D,D)$ provided ${\cal H}$ is replaced by  $ {\cal H} \rightarrow \tilde {\cal H}= ({\cal T}^{-1})^{t}  {\cal H} {\cal T}^{-1}$. Let us  remind that a $D\times D$ matrix ${\cal T}$ is an element of $O(D,D)$ if and only if:
\begin{equation}
{\cal T}^{t} \, \eta \, {\cal T} = \eta ,
\end{equation}
i.e. if it leaves the matrix $\eta$ invariant.  This definition implies that the $O(D,D)$ invariant metric by itself and the generalized metric ${\cal H}$ are elements of $O(D,D)$.

%In other words, $H$ is invariant under the above mentioned transformation if ${\cal H}$ is an element of $O(D,D)$.

%The inverse ${\cal H}^{-1}$ of the generalized metric  %${\cal T} \in O(D,D)$ 
%satisfies:
%\begin{eqnarray}
%{\cal H}^{-1} = \eta \,  {\cal H} \, \eta  \,\,\, .
%\end{eqnarray}

Alternatively, one can get the $O(D,D)$ invariance of $H$ also through the following considerations \cite{Rennecke}.
In terms of the generalized vector $A_P$, the constraints  coming from the vanishing of the two-dimensional energy-momentum tensor $T_{\alpha \beta}=0$. i.e.  from  the equation of motion of the world-sheet metric $h$
\begin{eqnarray}
G_{ab} \,  (\partial_{0} X^{a} \partial_{0} X^{b} + \partial_{1} X^a \partial_1X^{b} ) & =  & 0  \nonumber \\
G_{ab} \,\partial_{0} {X}^{a} \partial_{1} X^{b} & = & 0   \label{constraints}
\end{eqnarray} 
 can be  respectively rewritten as:
\begin{eqnarray}
A^T_P \,{\cal H}  \,A_P  =  0  \,\,\,\,\,\,\, ; \,\,\,\,\,\,\,\, 
A^T_P \, \eta \, A_P  =  0. 
\end{eqnarray}
The first constraint sets the Hamiltonian density to zero, while
the second one completely determines the string dynamics.  All the admissible generalized vectors satisfying the second constraint are related by an $O(D,D)$ transformation via $\tilde A_P=\mathcal{T}A_P$. In order to satisfy also the first constraint,  a compensating $O(D,D)$ transformation $\mathcal{T}^{-1}: {\mathcal H}\rightarrow ({\mathcal{T}^{-1}})^t  {\mathcal H} \mathcal{T}^{-1}  $ has to be applied to the generalized metric ${\cal H}$.  %In particular, the Hamiltonian density ${\cal H}$ is invariant under the interchange $\partial_{\sigma} X \leftrightarrow 2 \pi \alpha' P$ . One can think of this transformation as coming from a more fundamental symmetry of ${\cal H}$ under the exchange $\tau \leftrightarrow \sigma$, that seems to be at the root of T-duality.

In the presence of constant background $(G,B)$ along $d$ directions labelled by $a,b = (1,\dots, d)$ or, equivalently, in the presence of $d$ toroidally compactifed dimensions,  the equations of motion for the string coordinates  represent  a set of conservation laws of currents, locally defined on the world-sheet \cite{Maharana}:
\begin{equation}
\partial_{\alpha}J_a^{\alpha}=0,
\end{equation}
with 
\begin{equation}
J_a^{\alpha}=h^{\alpha \beta} G_{a b}\partial_{\beta}X^b+ \epsilon^{\alpha \beta} B_{ab} \partial_{\beta}X^b  \, . \label{currents}
\end{equation} 
Eq. (\ref{currents}) provides the following definition of the dual coordinate of $X^{a}$,  denoted by $\tilde{X}_a$ :
\begin{equation}
J_a^{\alpha} \equiv \epsilon^{\alpha \beta} \partial_{\beta} \tilde{X}_a \, .
\end{equation}
 By comparing the transformed metric $\tilde{ \mathcal{H} }\equiv \mathcal{H}^{-1}$ with $\mathcal{H}$ we may re-interpret its entries in terms of new fields $(\tilde{G}, \tilde{B})$ whose expression in terms of the previous ones can be easily read off. Thus, the initial Polyakov action $S$ defines a dual action $\tilde{S}$ which is a functional  of the constant dual $(\tilde{G}, \tilde{B})$-background. The relations:
\begin{equation}
\tilde{G}=\left(G-B G^{-1}B \right)^{-1}, \quad \tilde{B}=-G^{-1}B \tilde{G}
\end{equation}
are the so called   Buscher rules  \cite{Buscher1987,Buscher1988}.
Since $S$ and $\tilde{S}$ describe the evolution of the same string theory, they are dual to each other.

In terms of the generalized vector 
\begin{equation}
\chi \equiv \left( \begin{array}{c}
X  \\
  \tilde{X}
   \end{array} \right) 
\end{equation}
the equations of motion for $X$ and their dual analogues  have been shown in ref.s  \cite{Duff1990, Hull2005} to be recast into  a single $O(d,d)$-invariant equation:
\begin{equation}
\mathcal{H} \partial_{\alpha} \chi= \eta \, \epsilon_{\alpha \beta} \, \partial^{\beta}\chi
\end{equation}
involving both the generalized metric ${\cal H}$ and the $O(d,d)$-invariant metric $\eta$ previously defined. 

In particular, if the closed string coordinates are defined on a $d$-dimensional torus $T^d$, the dual coordinates will satisfy the same periodicity conditions and then $O(d,d) \rightarrow O(d,d;\mathbb{Z})$ becomes an exact symmetry: it constitutes what is called \textit{Abelian T-duality} \cite{Giveon1994} \cite{Alvarez1995}.

This has suggested since long \cite{ Duff1990, Hull2005, Tseytlin1990,  Tseytlin1991, Siegel1993, Lee2014} to look for a manifestly T-dual invariant formulation of string theory that  has to be based on a doubling of the string coordinates in the target space, since it requires the introduction of both the coordinates $X^a$ and the dual ones $\tilde{X}_a$. The main goal of this new action would be to explore more closely aspects of string geometry, hence of string gravity.

From a manifestly T-dual invariant two-dimensional string world-sheet, Double Field Theory (DFT) \cite{Hull2009} should emerge out as a low-energy limit. DFT developed as a way to encompass the Abelian T-duality in field theory and \textit{Doubled Geometry} underlies it. In DFT, diffeomorphisms rely on an $O(d,d)$ structure defined on the tangent space of a doubled torus $\mathcal{T}^{2d}$. A section condition has then to be imposed for halving the $2d$ coordinates. There is a vast literature concerning DFT  [18-33] %\cite{Aldazabal, Hohm2010, Blumenhagen2015a, Blumenhagen2015, Hull2009a, Nibbelink2013, Angelis2014, Copland2012, Park2013, Berman2008, Ma2015, Berman2015, Hassler2018, Pezzella2015, Pezzella2015a, Bandos2015} 
including topological aspects and its description on group manifolds.

Abelian T-duality requires the two dual target spaces to have Abelian isometry groups, but starting from the work in ref. \cite{Ossa1993}, it was clear that a non-Abelian generalization is possible, where one of the two isometry groups is non-Abelian. Recently, a further generalization has attracted growing interest, which is the so-called \textit{Poisson-Lie T-duality}  \cite{Klimcik1996a,Klimcik1996,Alekseev1996, Alekseev1994, Lledo1998}. Differently from the other two types, this latter does not require the presence of isometry groups. There are many reasons to consider this more general duality: at the classical level, a lot of models are concerned with T-duality without isometry and at the quantum level there arise possibilities of existence of non-trivial discrete symmetries relating a weak coupling regime of one quantum field theory to the strong coupling regime of the dual theory. Last but not least, it has  paved the way to new perspectives in string theory in non-geometric backgrounds \cite{Klimcik1997,Klimcik1996b}.

In particular, Poisson-Lie T-duality relies on the concept of \textit{Drinfel'd double} \cite{Drinfelextquotesingled1988}. As already remarked, the standard Abelian T-duality refers to the presence of Abelian isometries $U(1)^d$ in both the dual sigma models and they can be composed into $U(1)^{2d}$ that provides the simplest example of a Drinfel'd Double, i.e. a Lie group $D$ whose Lie algebra $\mathfrak{d}$ can be decomposed into a pair of maximally isotropic subalgebras with respect to a non-degenerate invariant bilinear form on $\mathfrak{d}$. %\cite{Drinfelextquotesingled1988}. 
We review the basic aspects of Drinfel'd doubles and its relation to T-dualities in Sect. \ref{secdrinfpoissonlie}.

 A classification of T-dualities can be given according to the types of underlying Drinfel'd doubles \cite{Klimcik1996a,Klimcik1996}:
\begin{itemize}
\item Abelian doubles correspond to the standard Abelian T-duality. The Drinfel'd double is Abelian, with Lie algebra given by the direct sum $\mathfrak{d}=\mathfrak{g}\, \oplus \, \tilde{\mathfrak{g}}$ of the Abelian algebra $\mathfrak{g}$ and its dual;
\item semi-Abelian doubles ($\mathfrak{d}=\mathfrak{g}\, \dot{\oplus} \, \tilde{\mathfrak{g}}$ with $\tilde{\mathfrak{g}}$ Abelian and $\dot{\oplus}$ indicating the semi-direct sum)  correspond to non-Abelian T-duality or, equivalently, 
they correspond to the standard non-Abelian T-duality between a  $\sigma$-model with a $G$-target, isometric with respect to  the Lie group $G$ having $\mathfrak{g}$ as its Lie algebra,  and a non-isometric $\sigma$-model with the target $\tilde{\mathfrak g}$ viewed as the Abelian group;
\item non-Abelian doubles (all the others) correspond to Poisson-Lie T-duality where no isometries hold for either of the two dual
models.
\end{itemize}

The Double Field Theory framework has also been considered in connection with \textit{Generalized Geometry} (GG) \cite{Hitchin2003,Hitchin2010,Gualtieri2007}, which has consequently arisen as a mean to \textit{geometrize} duality symmetries. GG is based on replacing the tangent bundle $TM$ of a manifold $M$ with a  kind of Whitney sum $TM \oplus T^* M$, a bundle with the same base space but fibers given by the direct sum of tangent and cotangent spaces, and the Lie brackets on the sections of $TM$ by the so-called \textit{Courant brackets}, involving vector fields and one-forms. The Courant bracket of two sections of $TM \oplus T^* M$ is defined as
\begin{equation}
[X+\xi, Y+ \eta]=[X,Y]+\mathcal{L}_X \eta-\mathcal{L}_Y \xi-\frac{1}{2} d\left(i_X \eta-i_Y \xi \right),
\end{equation}
where $X, Y$ are tangent vectors and $\xi, \eta$ cotangent vectors, such that $X+\xi$ and $Y+\eta$ are elements of the fibers $T \oplus T^*$. The important point of such bracket is that it commutes with the action of a closed $2$-form $B$ \cite{Hitchin2010}. Furthermore, it is easy to note that both this bracket and the inner product naturally defined on the generalized bundle $(x+\xi, X+\xi)=i_X \xi$ are invariant under diffeomorphisms of the underlying manifold $M$. This, together with the fact that a global closed $2$-form $B$ will also preserve both the inner product and Courant bracket, means an overall action of the semi-direct product of diffeomorphisms and closed $2$-forms. % $\mboxt{Diff}(M) \ltimes  \Lambda^2(\mbox{M})$.
 This formal setting has attracted interest in relation to DFT because it takes into account in a unified fashion vector fields, which generate diffeomorphisms for the $G_{ij}$ field, and one-forms, generating diffeomorphisms for the the $B_{ij}$ field.

Therefore, it seems relevant to analyze more deeply the geometrical
structure of (Abelian, non-Abelian, Poisson-Lie) T-dualities and
their relations with Generalized Geometry and/or Doubled
Geometry.

In this work we review the results presented in \cite{Marotta2018}, where a simple mechanical system has been considered: the three-dimensional isotropic rigid rotator (IRR), thought of as a $(0+1)$-field theory.

A remarkable property of this model is that its dynamics exhibits Poisson-Lie symmetries  \cite{MARMO1995,MarmoIbort} when described in the Hamiltonian approach, by replacing the cotangent space of $SU(2)$ with the group $SL(2,\mathbb{C})$ which plays the role of the alternative phase space of the model. The result is consistent with the two spaces, $T^*SU(2)$ and $SL(2,\mathbb{C})$, being symplectomorphic \cite{MarmoIbort}. Let us remark here that the concept of Poisson-Lie symmetry, which concerns a single dynamical model, can be stated independently and it is indeed a pre-requisite   for  Poisson-Lie duality, which requires instead two dynamical systems with different carrier spaces to be formulated. 
\\
With this distinction in mind, in \cite{Marotta2018} the IRR model was considered under yet another   point of view, the goal being to introduce a  model on the dual group of $SU(2)$, with the aim of exhibiting a  Poisson-Lie dual system, according to the definition given above. 
\\
It turned out that the model on the dual group doesn't describe the same dynamics, though paving the way to a field theory generalization of the whole construction, which describes the Principal Chiral Model of $SU(2)$ and its dual partner as Poisson-Lie duals \cite{Marotta2019}. The IRR is thus to be conceived as a toy model, where the key features of Poisson-Lie symmetries and Poisson-Lie duality can be clearly understood, although it should be stressed that its dynamics is not invariant under such transformations. 

After defining the dual model,  a parent action on the Drinfel'd double of $SU(2)$ is introduced, containing a number of degrees of freedom which is doubled with respect to the original one and from which both this latter  and its dual can be recovered by a suitable gauging. The geometric structures that appear can then be understood in terms of Generalized Geometry and/or Doubled Geometry.

\section{Drinfel'd Doubles and Poisson-Lie T-Duality}
\label{secdrinfpoissonlie}

In this section we shortly review the mathematical setting of Drinfel'd doubles and Poisson-Lie T-duality, with particular focus on $\mathfrak{sl}(2,\mathbb{C})$ as a double Lie algebra, which is needed in order to explore the relation between Drinfel'd doubles, Double Geometry and Generalized Geometry by introducing a new formulation of the IRR on $\mathfrak{sl}(2,\mathbb{C})$ with manifest symmetry under duality. More information on these topics can be found in  ref.s \cite{Plauschinn2018,Klimcik1996,Marotta2018,Marotta2018a,Hassler2017}, which are going to be closely followed in this section.

\subsection{Drinfel'd double structure of $\mathfrak{sl}(2,\mathbb{C})$}

A \textit{Drinfel'd double} $D$ is an even-dimensional Lie group whose Lie algebra $\mathfrak{d}$ can be decomposed into a pair of maximally isotropic subalgebras, $\mathfrak{g}$ and $\tilde{\mathfrak{g}}$, with respect to a non-degenerate invariant bilinear form on $\mathfrak{d}$.
About the terminology, for a subspace to be isotropic it means that the bilinear pairing of any two of its elements vanishes, while maximal refers to the fact that such subspace cannot be enlarged preserving the isotropy condition.

The associated Lie algebra is denoted by $\mathfrak{d}=\mathfrak{g} \bowtie \tilde{\mathfrak{g}}$ emphasizing the symmetric construction, and Lie brackets are given by
\begin{equation}
[T_i, T_j ] = {f_{ij}}^k T_k \,\,\,\,\,; \,\,\,\,\,
[\tilde{T}^i, \tilde{T}^j]  =  {g^{ij}}_k \tilde{T}^k \,\,\,\,\, ; \,\,\,\,\,   \label{drinfbrack}
[T_i, \tilde{T}^j]  =  {f_{ki}}^j\tilde{T}^k-{g^{kj}}_i T_k,
\end{equation}
being $T_i$ and $\tilde{T}^i$, with $i=1,2, \dots,  d$, respectively the generators of $\mathfrak{g}$ and $\tilde{\mathfrak{g}}$, both $d$-dimensional.  Let us notice that these Lie brackets emerge in the description of  the gauge algebra of string compactification on a Poisson-Lie background \cite{Reid-Edwards2010}. 

The two Lie subalgebras, when taken together,  define a Lie bialgebra $(\mathfrak{g},\tilde{\mathfrak{g}})$ and generate respectively the two Lie subgroups $G$ and $\tilde{G}$, such that $D=G \bowtie \tilde{G}$ (with an abuse of notation  we use the same symbol for the  algebra and group composition). In this picture it is clear what dual means: $G$ and $\tilde{G}$ are dual Lie groups in the sense that they are dual partners with respect to this decomposition. Note that since the role of the two subalgebras can be intercharged, $(\tilde{\mathfrak{g}},\mathfrak{g})$ is a Lie bialgebra as well. By construction, also $ \tilde{\mathfrak{d}}$ is a Lie bialgebra, and in particular $\left(\mathfrak{d}, \tilde{\mathfrak{d}} \right)$ is called the \textit{double} of $\left(\mathfrak{g}, \tilde{\mathfrak{g}} \right)$ \cite{Tian-Shansky1993}. 

In general, a particular splitting of the Drinfel'd double in terms of maximally isotropic subspaces is called a \textit{polarization}, and when the subspaces involved in the polarization close as subalgebras, the triple $(\mathfrak{d}, \mathfrak{g}, \tilde{\mathfrak{g}})$ is called a \textit{Manin triple} \cite{Drinfelextquotesingled1988,DRINFELextquotesingleD1990,Liu1997}. In this particular case, a natural $O(d,d)$ structure appears as duality pairing between the two subalgebras. In fact, the duality pairing between $\mathfrak{g}$ and $\tilde{\mathfrak{g}}$ with respect to which the isotropy follows is an invariant $O(d,d)$ metric $\eta$ with:
\begin{equation}
\label{pairing1}
\eta\left(T_i, T_j \right)=0, \quad \eta\left(\tilde{T}^i, \tilde{T}^j \right)=0,
\end{equation}
and 
\begin{equation}
\label{pairing2}
\eta\left(T_i, \tilde{T}^j \right)=\delta_i^j, \quad \eta\left(\tilde{T}^i, T_j \right)=\delta^i_j.
\end{equation}
The set of polarizations 
%the Manin triples 
corresponding to a given Drinfel'd double plays the role of the modular space of sigma models which, when related to Manin triples,  are mutually connected by Poisson-Lie T-duality. In the Abelian case the Drinfel'd double is $U(1)^{2d}$ and its modular space is nothing but $O(d,d,\mathbb{Z})$. Because of the symmetric construction and the symmetric role of the two subgroups, the modular space in question has always at least two points, originating from $G/\tilde{G}$ and the quotient obtained exchanging the groups \cite{Klimcik1996}.
Furthermore, it can be shown that a \textit{para-Hermitian} structure can be defined on $D$, induced by the Manin triple polarization $\mathfrak{d}=\mathfrak{g} \bowtie \tilde{\mathfrak{g}}$ \cite{Marotta2018,Marotta2018a}. 

For the purpose of this work, we will focus on the Drinfel'd double structure of $\mathfrak{d}=\mathfrak{sl}(2,\mathbb{C})$. 

It is known that the Lie algebra $\mathfrak{sl}(2,\mathbb{C)}$ can be regarded as a real form of the complex Lie algebra $\mathfrak{sl}(2)$. Indeed, $\mathfrak{sl}(2)$ is defined by the Lie brackets
\begin{eqnarray}
[t_1, t_2]   =  t_3  \,\,\,\,\, ; \,\,\,\,\,  [t_2, t_3] = 2t_2 \,\, \,\,\, ; \,\,\,\,\, 
[t_3, t_1]  =  2 t_1,
\end{eqnarray}
being 
\begin{equation}
t_1= \left( \begin{array}{cc}
  0 & 1 \\
  0 & 0
 \end{array} \right) ; \quad t_2= \left( \begin{array}{cc}
  0 & 0 \\
  1 & 0
 \end{array} \right) ; \quad t_3= \left( \begin{array}{cc}
  1 & 0 \\
  0 & -1
 \end{array}  \right) 
\end{equation}
its generators. However, by taking complex linear combinations of the $\mathfrak{sl}(2)$ generators given by
\begin{equation}
\label{su2basis}
e_1=\frac{1}{2}\left(t_1+t_2 \right)=\frac{\sigma_1}{2}, \quad e_2=\frac{i}{2}\left(t_2-t_1 \right)=\frac{\sigma_2}{2}, \quad e_3=\frac{1}{2}t_3=\frac{\sigma_3}{2},
\end{equation}
\begin{equation}
b_i=i e_i, \quad i=1,2,3
\end{equation}
the real algebra $\mathfrak{sl}(2,\mathbb{C})$ is recovered with its Lie brackets
\begin{eqnarray}
\label{liesl2c}
 [e_i, e_j]  =  i {\epsilon_{ij}}^k e_k \,\,\,\,\, ; \,\,\,\,\,
[e_i, b_j]  =  i {\epsilon_{ij}}^k b_k \,\,\,\,\,, ; \,\,\,\,\,
[b_i, b_j]  =  -i {\epsilon_{ij}}^k e_k.
\end{eqnarray}
From (\ref{liesl2c}) it is clear that the $e_i$'s are the generators of the $\mathfrak{su}(2)$ subalgebra.

One can also consider the dual vector space $\mathfrak{su}(2)^*$ by introducing a basis of vectors $\{\tilde{e}^j\}$ dual to $\{e_i \}$ in (\ref{su2basis}):
\begin{equation}
\label{sbgen}
\tilde{e}^i=\delta^{ij}\left(b_j+ {\epsilon^k}_{j3} \, e_k \right).
\end{equation}
This is indeed dual with respect to the natural scalar product (the Killing-Cartan form) defined in $\mathfrak{sl}(2,\mathbb{C})$ as
\begin{equation}
\label{sp1}
\langle v, w \rangle=2 \mbox{Im}\left[\mbox{Tr}(vw) \right], \quad \forall \, v,w, \in \mathfrak{sl}(2,\mathbb{C}),
\end{equation}
as can be easily seen by showing that
\begin{equation}
\langle \tilde{e}^i, e_j \rangle=2 \mbox{Im}\left[\mbox{Tr}(\tilde{e}^i e_j) \right]=\delta^i_j.
\end{equation}
The dual vector space $\mathfrak{su}(2)^*$ is in turn the Lie algebra $\mathfrak{sb}(2,\mathbb{C})$, with the following brackets:
\begin{equation}
\label{liesb}
[\tilde{e}^i, \tilde{e}^j]=i {f^{ij}}_k \tilde{e}^k,
\end{equation}
with $\mathfrak{sb}(2,\mathbb{C})$ structure constants ${f^{ij}}_k=\epsilon^{ijl} \epsilon_{l3k}$.

Furthermore, the Lie bracket
\begin{equation}
\label{liesusb}
[\tilde{e}^i, e_j]=i {\epsilon^i}_{jk} \tilde{e}^k+i {f^{ki}}_j e_k
\end{equation}
shows that each subalgebra acts on the other one non-trivially, by co-adjoint action.

It is important to note that both $\mathfrak{su}(2)$ and $\mathfrak{sb}(2,\mathbb{C})$ are maximally isotropic with respect to the scalar product (\ref{sp1}):
\begin{equation}
\langle e_i, e_j \rangle=0, \quad \langle \tilde{e}^i, \tilde{e}^j \rangle=0,
\end{equation}
therefore, $(\mathfrak{sl}(2,\mathbb{C}), \mathfrak{su}(2), \mathfrak{sb}(2,\mathbb{C}))$ is a Manin triple with respect to it, and $\mathfrak{sl}(2,\mathbb{C})$ can be understood as a Drinfel'd double with polarization $\mathfrak{sl}(2,\mathbb{C})=\mathfrak{su}(2) \bowtie \mathfrak{sb}(2,\mathbb{C})$. $SU(2)$ and $SB(2,\mathbb{C})$ are then dual groups, in the sense that they are dual partners with respect to this particular splitting. Observe that the Lie brackets of (\ref{liesl2c}) (for the $e_i$), (\ref{liesb}) and (\ref{liesusb}) have exactly the form as (\ref{drinfbrack}).

There is also another non-degenerate invariant scalar product which can be defined on $\mathfrak{sl}(2,\mathbb{C})$ as:
\begin{equation}
\label{sp2}
(v,w)=2 \mbox{Re}\left[\mbox{Tr}\left(vw \right) \right], \quad \forall \, v,w \in \mathfrak{sl}(2,\mathbb{C}).
\end{equation}
$\mathfrak{su}(2)$ and $\mathfrak{sb}(2,\mathbb{C})$ are no longer isotropic subspaces with respect to this scalar product, in fact, for the basis elements:
\begin{equation}
(e_i, e_j)=\delta_{ij}, \quad (b_i, b_j)=-\delta_{ij}, \quad (e_i, b_j)=0.
\end{equation}
Note that this does not give rise to a positive-definite metric. However, on denoting by $C_+$, $C_-$ respectively the two subspaces spanned by $\{e_i \}$ and $\{ b_i\}$, the splitting $\mathfrak{sl}(2,\mathbb{C})=C_+ \oplus C_-$ (which is not a Manin triple polarization) defines a positive definite metric $\mathcal{H}$ on $\mathfrak{sl}(2,\mathbb{C})$ as follows:
\begin{equation}
\mathcal{H}=\left( , \right)_{C_+}-\left( , \right)_{C_-}.
\end{equation}
This is a Riemannian metric and we denote it with the symbol $\left( \left(	\, ,\,  \right) \right)$. In particular:
\begin{equation}
\label{riemannianmetric}
\left( \left(e_i, e_j \right) \right) \equiv  \left(e_i, e_j \right), \quad \left( \left( b_i, b_j \right) \right) \equiv -\left(b_i, b_j \right), \quad \left( \left(e_i, b_j \right) \right) \equiv \left(e_i ,b_j\right)=0.
\end{equation}
In Sect. \ref{secdoubled} we will see that in doubled notation $e_I= \left( \begin{array}{c}
  e_i  \\
  \tilde{e}^i
   \end{array} \right),$ with $e_i \in \mathfrak{su}(2)$ and $\tilde{e}^i \in \mathfrak{sb}(2,\mathbb{C})$, the scalar product in (\ref{sp1}) defines precisely the $O(d,d)$ (in this case $d=3)$ invariant metric  while the Riemannian scalar product in (\ref{riemannianmetric}) defines a pseudo-orthogonal $O(d,d)$ matrix that will correspond to the generalized metric. 
   
As a further interesting remark, the Riemannian structure $\mathcal{H}$ can be mathematically formalized in a way which clarifies its role in connection with Generalized Geometry \cite{Gualtieri2007}\cite{Freidel2017}: it can be related to the structure of para-Hermitian manifold of $SL(2,C)$ and therefore generalized to even-dimensional real manifolds which are not Lie groups.

\subsection{Poisson-Lie T-duality}
\label{secpoisslieduality}

As announced in the introduction, in order to formulate the invariance under Poisson-Lie T-duality as a generalization of Abelian and semi-Abelian T-dualities, we need to state the concept of Poisson-Lie symmetry. 

%\noindentation
{\bf Definition 1} A coordinate  transformation associated to the Lie group $G$ is a symmetry of the dynamics if it leaves the equations of motion unchanged in form. 

If it is also a symmetry for the auxiliary geometric structures which are relevant to the chosen formulation (symplectic form,  Poisson brackets and Hamiltonian in the Hamiltonian approach, functional action and Lagrangian in the Lagrangian approach), namely an isometry, then the symmetry yields constants of motion, which are associated to conserved N\"other currents.

%\noindentation
{\bf Definition 2} If a given symmetry of the dynamics under the Lie group $G$ is not a symmetry for the auxiliary geometric structures, that is not an isometry,  but the violation is governed by the Maurer-Cartan structure equation of the  Drinfel'd dual  group $\tilde G$ associated to the group $G$, the symmetry of the dynamics is a Poisson-Lie symmetry.

Let us be more specific. To be definite, let us choose the Hamiltonian approach. Let us assume that the dynamics of the system under consideration is invariant with respect the action of a given Lie group $G$, namely, Hamilton equations of motion are unchanged. 
These are, for a certain observable $f(q,p)$, the following:
\begin{equation} \dot f = \Lambda (df, dH) \label{hameq}
\end{equation}
with $\Lambda$ the Poisson tensor and $H$ the Hamiltonian. They are certainly invariant under a given Lie group action with infinitesimal generator $X_a$ if both $\Lambda$ (or the symplectic form $\omega$), and $H$ are invariant
\begin{equation}
\mathcal{L}_{{X_a}} \Lambda = 0,\; \;\; \mathcal{L}_{{X_a}} \omega= 0 ; \;\;\;\; 
\mathcal{L}_{{X_a}} H= 0
\end{equation}
or, in other words, we have conserved quantities defined by 
\begin{equation}
i_{{X_a}}\omega = d h_a
\end{equation}
but this is only a sufficient condition. In order for the dynamics to be invariant we might modify both the Poisson brackets and the Hamiltonian in such a way that their Lie derivative is non-zero, but  the eq. (\ref {hameq})  does not change. In terms of the symplectic form this implies that $\theta^a= i_{{X_a}}\omega$ is not an exact form anymore, namely $d\theta^a\ne 0 $ and it does not define a constant of motion.
If $\theta^a$ is such that
\begin{equation}
d \theta^a = -\frac{1}{2} { \tilde{f}^{a}}_{\, \, \, \, bc} \theta^b\wedge \theta^c
\end{equation}
with ${ \tilde{f}^{a}}_{\, \, \, \, bc}$ structure constants of a Lie group $\tilde G$, which can be determined algebraically by requiring the product $G\bowtie \tilde G$ to be the Drinfel'd double of $G$, the symmetry of the dynamics under the action of $G$ is a Poisson-Lie symmetry and the one-forms associated to the infinitesimal generators of the $G$ symmetry through the symplectic form are Maurer-Cartan one-forms of the dual group $\tilde G$.

Once a Poisson-Lie symmetry is found  for a given dynamical model, under the action of a Lie group $G$, one can construct a model with a different target space, which is acted on by the dual group $\tilde G$. If this model possesses a Poisson-Lie symmetry as well, namely the one forms associated to the infinitesimal generators of $\tilde{G}$ through its symplectic form $\tilde \omega$ are Maurer-Cartan one-forms for $G$, the two models are dual to each other and the symmetry under duality is what is called an invariance under Poisson-Lie duality. 

In order to understand this picture on an interesting model, and reconnect to definitions commonly used in the literature, which mainly rely on the Lagrangian approach, let us consider the bosonic string sigma model in a target space $M$ with $(G,B)$ background.

We demand the target space to admit a free action of a Lie group $G$ from the right. The action of the theory in eq. (\ref{action}) can be rewritten by introducing complex coordinates $z$ and $\bar{z}$ on the world-sheet, in a convenient form as:
\begin{equation}
S=  \frac{1}{4 \pi \alpha'} \int dz d \bar{z} \, E_{ij}  \partial X ^i \bar{\partial}X^j, 
\end{equation}
with $E_{ij}=G_{ij}+B_{ij}$ a matrix combining the metric tensor on the target space and the antisymmetric Kalb-Ramond $B$ field. By varying the action with respect to the coordinates, with  $\delta X^a=V^{a}_{\,\,\,b} \delta \epsilon^b$,  $V^a_{\,\,\, b}$ the left-invariant vector fields on $M$ and $ \epsilon^b$ infinitesimal parameters of the transformation, one gets the Noether currents:
\begin{equation}
J_a=V_a^i \left( E_{ij} \bar{\partial} X^j d\bar{z} -E_{ji} \partial X^j dz\right).
\end{equation}

This can be proved performing explicitly the variation of the action, obtaining:
\begin{equation}
4\pi \alpha' \delta S= \int dz d\bar{z} \, \mathcal{L}_{V_a} (E_{ij}) \partial X^i \bar{\partial}X^j \epsilon^a-\int dz d\bar{z}\left[\partial (V_a^i E_{ij}\bar{\partial}X^j)+\bar{\partial}(V_a^i E_{ji} \partial X^j) \right]\epsilon^a,
\end{equation}
where $\mathcal{L}_{V_a}$ denotes the Lie derivative along the vector field $V_a$, and realizing that 
\begin{equation}
dz d\bar{z}\left[\partial (V_a^i E_{ij}\bar{\partial}X^j)+\bar{\partial}(V_a^i E_{ji} \partial X^j)\right]=d\left(  V_a^i E_{ij} \bar{\partial} X^j d\bar{z} -V_a^i E_{ji} \partial X^j dz \right),
\end{equation}
so we have
\begin{equation}
4\pi \alpha' \delta S= \int dz d\bar{z} \, \mathcal{L}_{V_a} (E_{ij}) \partial X^i \bar{\partial}X^j \epsilon^a-\int dJ_a \, \epsilon^a.
\end{equation}

In the standard T-duality approach,  one would have the current one-forms $J_a$ to be closed, which follows from the requirement that the Lie derivative along $V$ acting on $G$ and $B$ vanishes, namely, besides an   invariance of the dynamics, we have  also an invariance of the geometric structures. Therefore,  the left-invariant vector fields $V_a$ have to be Killing vectors, and T-duality is along a direction of isometry. This is the standard picture where isometry is the founding ingredient. However, if we are in the presence of a symmetry of the dynamics which is not an isometry, but  the currents obey  the following on-shell integrability condition (the  Maurer-Cartan equation advocated above)
\begin{equation}
\label{integrcurrent}
dJ_a-\frac{1}{2} {\tilde{f}^{\,\,\,bc}}_a J_b \wedge J_c=0,
\end{equation}
with ${\tilde{f}^{bc}}_a$ structure constants of a certain Lie algebra,  the $V_a$'s do not correspond to isometries. This can also be seen by using both the integrability condition and the fact that on-shell the variation of the action vanishes:
\begin{equation}
\label{actvan}
\int dz d\bar{z} \, \mathcal{L}_{V_a} (E_{ij}) \partial X^i \bar{\partial}X^j \epsilon^a=\int \frac{1}{2} {\tilde{f}^{\,\,\,bc}}_a J_b \wedge J_c \, \epsilon^a.
\end{equation}
It is straightforward to obtain
\begin{equation}
J_b 	\wedge J_c=-2 V_b^m V_c^l E_{nm} E_{lk}\partial X^n \bar{\partial}X^k dz d\bar{z},
\end{equation}
and putting this in  eq.( \ref{actvan}), one has:
\begin{equation}
\label{liederivativegeneralizedmatrix}
\mathcal{L}_{V_a} E_{ij}=-{\tilde{f}^{\,\,\,bc}}_a V_b^k V_c^l E_{ik} E_{lj}.
\end{equation}

If the currents are not closed but satisfy the integrability condition, it turns out that ${\tilde{f}_a^{\,\,bc}}$ are the structure constants of the Lie algebra $\tilde{\mathfrak{g}}$ corresponding to the dual Lie group $\tilde{G}$ (in the sense of Drinfel'd) and the theory is said to have a Poisson-Lie symmetry. 
Hence, in order to formulate Poisson-Lie T-duality, in contrast to the Abelian and non-Abelian versions,   isometries are not required. It is a more general version since the two latter versions are particular cases. In fact, if isometries are present, the dual Lie group is Abelian and $dJ_a=0$, which is a particular case of (\ref{integrcurrent}). Therefore, we understand that, formally, (\ref{integrcurrent}) is the Maurer-Cartan equation for the current one-forms on the dual group $\tilde{\mathfrak{g}}$. 

Being $V_a$ the left-invariant vector fields on the Lie group $G$, by using the commutators of the Lie derivatives along these fields, the Lie algebra $\mathfrak{g}$ of $G$ can be recovered:
\begin{equation}
\label{commutatorslivf}
\left[\mathcal{L}_{V_a}, \mathcal{L}_{V_b} \right]=\mathcal{L}_{{f_{ba}}^c V_c}.
\end{equation}
By combining the expression for the variation of $E_{ab}$ along the left-invariant vector fields (\ref{liederivativegeneralizedmatrix}) with the commutation relations (\ref{commutatorslivf}), a compatibility relation between the structure constants of the two Lie algebras is obtained:
\begin{equation}
{f_{ae}}^c \tilde{f}^{\,\,\, ed}_b+{f_{ae}}^d {\tilde{f}^{\,\,\, ce}}_b-{f_{be}}^c {\tilde{f}^{\,\,\, ed}}_a-{f_{be}}^d {\tilde{f}^{\,\,\, ce}}_a={f_{ab}}^e {\tilde{f}^{\,\,\, cd}}_e,
\end{equation}
which, considering $\tilde{\mathfrak{g}}$ as the dual vector space of $\mathfrak{g}$, is the Lie bialgebra structure.
It is possible to show that this compatibility relation requires these two algebras to be the maximally isotropic subalgebras of a Drinfel'd double $\mathfrak{d}=\mathfrak{g} \bowtie \tilde{\mathfrak{g}}$ with respect to a bilinear invariant pairing such as (\ref{pairing1}), (\ref{pairing2}). 
This is the reason why Drinfel'd doubles are the main objects in Poisson-Lie T-dualities.

Following the approach described above, in order to obtain Poisson-Lie T-dual models, one has to investigate the possibility of building a model on a target space which is acted upon by the group $\tilde G$ and study its symmetries. This is precisely the purpose of studying the IRR model and its dual partner on the Lie group $SB(2 \mathbb{C})$. 
%The dual sigma model with action $\tilde{S}=  \frac{1}{4 \pi \alpha'} \int dz d \bar{z} \, \tilde{E}_{ab}  \partial \tilde{X} ^a \bar{\partial}\tilde{X}^b$ can then be obtained after exchanging the two subgroups $G$ and $\tilde{G}$. The whole description can be lifted to the Drinfel'd double $D$ with a kind of Iwasawa decomposition $\gamma=\tilde{g} g$, with $\tilde{g} \in \tilde{\mathfrak{g}}$, $g \in \mathfrak{g}$, $\gamma \in D$ (or $\gamma=g \tilde{g}$) and the two dual models can be recovered by imposing appropriate gauging conditions. For our model we see this explicitly in Sect. \ref{secdoubled}. The dual models, when expressed in the Hamiltonian formalism, are related by a canonical transformation on the phase-space variables \cite{Klimcik1996a}.

\section{The Isotropic Rigid Rotator}
\label{secIRR}

In this section, following \cite{Marotta2018}, we consider an explicit and simple model: the three-dimensional isotropic rigid rotator, considered as a $(0+1)$-dimensional $SU(2)$ valued field theory. The model is simple but nonetheless it can give precious information, especially in view of its direct true field theory generalization that is the \textit{principal chiral model}, analysed in \cite{Marotta2019} having this simple model as a guide. This is not the only generalization whose dynamics can be captured by the rotator, such as the Wess-Zumino-Witten model (whose results are going to be detailed in a forthcoming paper \cite{WZW} or $(2+1)$-dimensional gravity in its first order formulation \cite{Witten1988} among many others. 

A suitable action for the IRR model is the following 
\begin{equation} \label{S0action}
S_0=\int_{\mathbb{R}}dt \, L_0=-\frac{1}{4}\int_{\mathbb{R}} \mbox{Tr}\left[\varphi^*\left(g^{-1}dg\right) \wedge *\varphi^*\left(g^{-1}dg\right) \right]=-\frac{1}{4}\int_{\mathbb{R}}dt \, \mbox{Tr}\left(g^{-1}\dot{g}\right)^2,
\end{equation}
with $\varphi \, : \, t \in \mathbb{R} \rightarrow g \in SU(2)$, $*$ the Hodge star operator on $\mathbb{R}$, defined such that $*dt=1$ and $\varphi^*$ denotes the pull-back map,  so that $\varphi^*\left(g^{-1}dg\right)=g^{-1}\partial_t g \, dt$ defines the pull-back of the Maurer-Cartan left-invariant one-form $g^{-1}dg$ on $\mathbb{R}$.\\
In particular, being $g^{-1}dg \in \Omega^1 \otimes \mathfrak{su}(2)$: $g^{-1} dg=i \alpha^k \sigma_k$, with $\sigma_k$ the Pauli matrices and $\alpha^k$ basic left-invariant one-forms. Here $\Omega^1$ denotes the space of one-forms on the group manifold.\\
The group manifold $SU(2)$ can be parametrized by the embedding in the ambient space $\mathbb{R}^4$ as follows:
\begin{equation}
g=2\left(y^0 e_0+i y^i e_i \right), \quad g \in SU(2),
\end{equation}
with $\left(y^0 \right)^2 + \sum_{i} \left(y^i \right)^2=1$, $e_0=\mathbb{I}/2$, $e_i=\sigma_i/2$.\\
One has then:
\begin{equation}
y^0=\langle e_0 | g \rangle=\mbox{Tr}\left(g e_0 \right), y^i=\langle e_i | g \rangle=-i \, \mbox{Tr}\left(ge_i \right), \quad i=1, 2, 3.
\end{equation}
On $SU(2)$ we have $g^{-1}=g^{\dagger}$, so that
\begin{eqnarray}
g^{-1}\dot{g} {} &= & \left(y^0 \mathbb{I}-iy^i\sigma_i \right) \left(\dot{y}^0 \mathbb{I}+i \dot{y}^j\sigma_j \right)= y^0 \dot{y}^0 \mathbb{I}+i y^0 \dot{y}^i \sigma_i-i y^i \dot{y}^0\sigma_i+y^i\dot{y}^j \sigma_i \sigma_j\\ & = & i\left(y^0 \dot{y}^i-y^i\dot{y}^0 + \epsilon^i_{\,\;jk}y^j \dot{y}^k\right)\sigma_i+\left(y^0 \dot{y}^0+y^i \dot{y}^i \right),
\end{eqnarray}
where we used the well known relation $\sigma_i \sigma_j=\delta_{ij}\mathbb{I}+i \epsilon_{ij}^{\, \, \;k} \sigma_k$.
Moreover, the last term appearing in the above equation is vanishing since $\left(y^0 \dot{y}^0+y^i \dot{y}^i \right)=\frac{1}{2}\frac{d}{dt}\left( \left(y^0 \right)^2 + \sum_{i} \left(y^i \right)^2\right)=0$.\\
This leads to
\begin{equation}
g^{-1}\dot{g}=i\left(y^0 \dot{y}^i-y^i\dot{y}^0 + \epsilon^i_{\, \;jk}y^j \dot{y}^k\right)\sigma_i=i \dot{Q}^i \sigma_i,
\end{equation}
defining the left generalized velocities $\dot{Q}^i \equiv \left(y^0 \dot{y}^i-y^i\dot{y}^0 + \epsilon^i_{\, \;jk}y^j \dot{y}^k\right)$, which allow us to write the Lagrangian as $L_0=\frac{1}{2}\dot{Q}^i \dot{Q}^j \delta_{ij}$. This can be seen as follows:
\begin{eqnarray}
L_0=-\frac{1}{4}\mbox{Tr}\left(g^{-1}\dot{g}\right)^2=-\frac{1}{4}\mbox{Tr}\left[\left(i\dot{Q}^i \sigma_i\right) \left(i \dot{Q}^j \sigma_j \right)\right]=\frac{1}{4}\mbox{Tr}\left[\dot{Q}^i \dot{Q}^j\left(\delta_{ij}\mathbb{I}+i \epsilon_{ij}^{\, \, \;k} \sigma_k \right) \right],
\end{eqnarray}
using the fact that the $\sigma$ matrices are traceless.

The Euler-Lagrangian equation of motion can be written in its intrinsic formulation \cite{Marmo1985}, especially relevant for non-invariant Lagrangians \footnote{This is not the case, but it will be useful for the dual model which we will show to have non-invariant Lagrangian.}, as:
\begin{equation}
\mathcal{L}_{\Gamma} \theta_L-dL_0=0,
\end{equation}
being $\mathcal{L}_{\Gamma}$ the Lie derivative along the $\Gamma=\frac{d}{dt}$ vector field and $\theta_L$ the Lagrangian one-form, which is given by $ \theta_L=\frac{\partial L}{\partial \dot Q^j} \alpha^j=
%-\frac{1}{4}\text{Tr}\left(g^{-1}\dot{g} g^{-1}dg\right)=-\frac{1}{4}\text{Tr}\left(-\dot{Q}^i \alpha^k \sigma_i \sigma_k\right)=
\frac{1}{2} \dot{Q}^i \alpha^j \delta_{ij}$.
%, where we used the usual relation for the product of Pauli matrices.

By projecting along the basic left-invariant vector fields $X_i$ (dual to the basic left-invariant one-forms $\alpha^i$), one obtains:
\begin{equation}
i_{X_i}\left(\mathcal{L}_{\Gamma} \theta_L-dL_0 \right)=0,
\end{equation}
where $i_{X_i}$ is the notation for the interior derivative along the vector field $X_i$. Since $\mathcal{L}_{\Gamma}$ and $i_{X_i}$ commute over the Lagrangian one-form \footnote{This is general: $i_X \mathcal{L}=i_X\left( i_X d+ d \, i_X\right)=i_X d \, i_X$ since $i_X$ is 2-nilpotent, while $\mathcal{L} i_X=\left(i_X d+ d \, i_X \right)i_X=i_X d \, i_X$. }, one gets:
\begin{equation}
\mathcal{L}_{\Gamma}\left(\frac{1}{2}\dot{Q}^j i_{X_i}\alpha^l \right)\delta_{jl}-\mathcal{L}_{X_i}L_0=0,
\end{equation}
where we have used the fact that $i_X d f = \mathcal{L}_X f$ for $f$ a function. 
%since the Lie derivative of a function $L_0$ with respect to a vector field $X_i$ is the same as the directional derivative $X_i(L_0)$, it is also the same as the contraction of the exterior derivative of $L_0$  with $X_i$: $i_{X_i} d L_0=\mathcal{L}_{X_i}L_0$.
Since $i_{X_i}\alpha^l=\delta_i^l$ and $\mathcal{L}_{X_i}L_0=\frac{1}{2}\dot{Q}^p \dot{Q}^q \epsilon_{ip}^{\, \; \, k} \delta_{qk}$, we are left with the equation of motion
\begin{equation}
\mathcal{L}_{\Gamma}\dot{Q}^j \delta_{ji}-\dot{Q}^p \dot{Q}^q \epsilon_{ip}^{\, \; \, k} \delta_{qk}=0,
\end{equation}
but the latter term is vanishing because it is the contraction of a symmetric and of an antisymmetric tensor, hence
\begin{equation}
\ddot{Q}^i=0, \quad i=1,2,3.
\end{equation}
Left momenta can be calculated as usual as:
\begin{equation}
I_i=\frac{\partial L_0}{\partial \dot{Q}^i}=\delta_{ij}\dot{Q}^j,
\end{equation}
and cotangent bundle coordinates can then be chosen to be $\left(Q^i, I_i \right)$.\\
The Legendre transform from $TSU(2)$ to $T^*SU(2)$ yields the Hamiltonian function:
\begin{equation}
H_0=\left[I_i \dot{Q}^i-L_0 \right]_{\dot{Q}^i=\delta^{ij}I_j}=\delta^{ij}I_i I_j-\frac{1}{2}\delta^{ij}I_j I_k \delta^{lk}\delta_{il}=\frac{1}{2}\delta^{ij}I_i I_j.
\end{equation}
By introducing the dual basis $\left\{e^{i^*} \right\}$ in the cotangent space, such that $\langle e^{i^*}|e_j \rangle=\delta^i_j$, one can consider the form 
\begin{equation}
I=-\frac{1}{2}iI_i e^{i^*}.
\end{equation}
In the first order formulation the action results to be
\begin{equation}
S=\int \theta  - \int dt H_0,
\end{equation}
where 
\begin{equation}
\theta = \langle I | g^{-1}dg \rangle=\langle -\frac{1}{2}iI_i e^{i^*} | 2i \alpha^k e_k \rangle=I_i  \alpha^k \delta^{i}_k
\end{equation}
is the canonical one-form, and reminding that the generators are defined as $e_i=\sigma_i/2$.
The symplectic form can then be obtained as
\begin{equation}
\omega=d\theta=dI_i \wedge \delta^i_j \alpha^j+I_i  \delta^i_j d\alpha^j=dI_i \wedge \delta^i_j \alpha^j+I_i \delta^i_j \epsilon^j_{\, \;lk} \alpha^l \wedge \alpha^k,
\end{equation}
where we also used the Maurer-Cartan equation $d\alpha^k=\epsilon^k_{ \, \;ij}\alpha^i \wedge \alpha^j$.
%\footnote{Maurer-Cartan equation in general can be derived as follows: since the Maurer-Cartan one-form is a Lie algebra valued one-form ($g^{-1}dg \in \Lambda^1 \otimes \mathfrak{g}$), for a generic group $G$ it can be written as $g^{-1}dg=\alpha^a T_a$, where $\alpha^a$ are the basic one-forms and $T_a$ denotes the generators of the Lie algebra $\mathfrak{g}$. We can then calculate $d\left(g^{-1}dg \right)=-g^{-1}dg \wedge g^{-1}dg=-\frac{1}{2}\alpha^a \wedge \alpha^b \left[T_a,T_b \right]=-\frac{1}{2}f_{ab}^{\, \, \; c} \alpha^a \wedge \alpha^b T_c$. At the same time $d\left(g^{-1}dg \right)=d\alpha^c T_c$, and the result follows by equating the two: $d\alpha^c=-\frac{1}{2}f_{ab}^{\, \, \; c}\alpha^a \wedge \alpha^b$.}
One can then calculate Poisson brackets by inverting $\omega$. The corresponding bi-vector field $\Lambda$ will be written in terms of the basis vector fields $\partial_{I_j}, X_j$ respectively spanning the fibers and the base manifold  of the cotangent bundle.
%be done with respect to the basic left-invariant vector fields $X_i$ instead of the $\del_{y^i}$ fields because these are redundant since they parametrize $\mathbb{R}^4$ embedding of $SU(2)$. 
Using the fact that $X_i \left(\alpha^j \right)=\delta_i^j$, $\frac{\partial}{\partial I_i}\left(dI_j\right)=\delta^i_j$, we have
\begin{equation}
\omega^{-1}=\Lambda=a_i^{\, \,j}\frac{\partial}{\partial I_i}\wedge X_j+b_{ij}\frac{\partial}{\partial I_i}\wedge\frac{\partial}{\partial I_j}+c^{ij}X_i \wedge X_j.
\end{equation}
By imposing the inverse condition one can easily see that $a_i^j=-\delta_i^j$, $b_{ij}=\epsilon_{ij}^{\, \; \, \;k} I_k$ and $c^{ij}=0$, so that %$\left\{\alpha^i,\alpha^j \right\}=c^{ij}=0 \Rightarrow 
$\left\{y^i,y^j \right\}=0$ and $\left\{I_i,I_j \right\}=\epsilon_{ij}^{\, \; \, \;k}I_k$. In order to calculate the $\left\{y^i,I_j \right\}$ bracket, one has to use  the expression of the left invariant vector fields in the chosen $\mathbb{R}^4$ parametrization.   To this, by recalling that  $\alpha^j= \frac{1}{2}\mbox{Tr} g^{-1} dg \sigma^j =y^0 dy^j-y^jdy^0+\epsilon_{lk}^{\, \, \; \, j}y^l dy^k$, and using the property  $X_i \left(\alpha^j \right)=\delta_i^j$ we get:
\begin{equation}
X_j=y^0 \frac{\partial}{\partial y^j}-y^j\frac{\partial}{\partial y^0}+\epsilon_{lj}^{\, \, \, k}y^l\frac{\partial}{\partial y^k}.
\end{equation}
implying
\begin{equation}
\left\{I_l,y^m \right\}=\Lambda (d I_l, dy^m)=-\delta_i^j\frac{\partial I_l}{\partial I_i}X_j(y^m)=-\delta_i^j\delta^i_l X_j(y^m),
\end{equation}
but since:
\begin{equation}
X_j(y^m)=y^0\frac{\partial}{\partial y^j}y^m-y^i\frac{\partial}{\partial y^0}y^m+\epsilon^s_{\, \,\, pj}y^p\frac{\partial}{\partial y^s}y^m=y^0 \delta^m_j+\delta^m_s \epsilon^s_{\, \,\, pj}y^p,
\end{equation}
one gets:
\begin{equation}
\left\{I_l,y^m \right\}=-\delta_i^j\delta^i_l X_j(y^m)=-\delta_i^j\delta^i_l\left(y^0 \delta^m_j+\delta^m_s \epsilon^s_{\, \,\, pj}y^p \right)=-y^0 \delta_l^m-\epsilon^m_{\, \,\, pj}y^p.
\end{equation}
From the Poisson brackets
\begin{eqnarray}
 \left\{y^i,y^j \right\} & = & 0 \\
\left\{I_i,I_j \right\} & = & \epsilon_{ij}^{\, \; \, \;k}I_k \\
\left\{y^i,I_j \right\} & = & \delta^i_j y^0+\epsilon^i_{\, \;jk}y^k \leftrightarrow \left\{g,I_j \right\}=2ige_j,
\end{eqnarray}
the Hamilton equations of motion can be derived:
\begin{eqnarray}
\dot{I}_i  =  \left\{I_i,H \right\}=0, \\ 
\dot{g}=\left\{g,H \right\}=-\delta^{ij}I_i \left\{I_j,g \right\}=2\delta^{ij}I_i i g e_j,
\end{eqnarray}
leading to 
\begin{equation}
g^{-1}\dot{g}=2i I_i \delta^{ij}e_j.
\end{equation}
These equations show that the fiber coordinates $I_i$, associated to the angular momentum components, are constants of motion as expected, while $g$ undergoes a uniform precession. In this case, since the Lagrangian and the Hamiltonian are invariant under both left and right action of the $SU(2)$ group, also the right momenta can be seen to be conserved as well, making the model super-integrable. This will not be the case for the dual model, as we show in Sect. \ref{secdualmodel}.

The fibers  of the tangent bundle $TSU(2)$ are, as vector space, $\mathfrak{su}(2) \simeq \mathbb{R}^3$, being $\dot{Q}^i$ vector fields components. 
The fibers of the cotangent bundle $T^* SU(2)$ are isomorphic to the dual Lie algebra $\mathfrak{su}(2)^*$. This, as a vector space, is again $\mathbb{R}^3$, but now $I_i$ are one-form components.

The carrier space of the Hamiltonian dynamics $T^*SU(2)$ is represented, as a group, by the semi-direct product of $SU(2)$ and the Abelian group $\mathbb{R}^3$, i.e. $T^* SU(2) \simeq SU(2) \ltimes \mathbb{R}^3$, with Lie algebra
\begin{eqnarray}
 \left[L_i,L_j \right] & = & i {\epsilon_{ij}}^k L_k \\ 
\left[T_i,T_j \right] & = & 0 \\
\left[L_i,T_j \right] &= & i {\epsilon_{ij}}^k T_k,
\end{eqnarray}
being $L_i$ the generators of the $SU(2)$ algebra and $T_i$ the generators of $\mathbb{R}^3$, which behave as vectors under $SU(2)$ rotations as can be seen from the last relation. 
The linearization of the Poisson structure at the identity of $SU(2)$ provides a Lie algebra structure over the dual algebra $\mathfrak{su}(2)^*$. Thus, the brackets $\{I_i ,I_j\}={\epsilon_{ij}}^k I_k$ are induced by the coadjoint action of the group $SU(2)$ on its dual algebra, hence the Poisson brackets governing the dynamics of the IRR are the Kirillov-Souriau-Konstant (KSK) brackets.

It has been shown in \cite{MARMO1995} that the carrier space of the dynamics of the IRR can be generalized to the semisimple Lie group $SL(2,\mathbb{C})$. This can be realized by deforming the Abelian subgroup $\mathbb{R}^3$ into the non-Abelian group $SB(2, \mathbb{C})$, which, we recall, is the Borel Lie subgroup of $2 \times 2$ upper triangular complex matrices with unit determinant. In particular, $SU(2)$ and $SB(2, \mathbb{C})$ constitute the pair with respect to which $SL(2,\mathbb{C})$ can be regarded as a Drinfel'd double. This means that the triple $\left(\mathfrak{sl}(2,\mathbb{C}),\mathfrak{su}(2), \mathfrak{sb}(2,\mathbb{C})\right)$ is a Manin triple with respect to the scalar product $\langle \cdot , \cdot \rangle$ in $\mathfrak{sl}\left(2, \mathbb{C} \right)$ defined in (\ref{sp1}), as we have discussed in Sect. \ref{secdrinfpoissonlie}. In the next section we discuss a new model, which will be referred to as dual,  in the sense that it is the analogue of the IRR but modeled on the dual group $SB(2, \mathbb{C})$.

\section{The Dual Model}
\label{secdualmodel}

 As carrier space for the dynamics of the dual model in the Lagrangian (respectively Hamiltonian) formulation one can choose  the tangent (respectively cotangent) bundle of the group $SB(2,\mathbb{C})$.
 The latter  is  the dual Lie group of $SU(2)$ because they are dual partners in a particular polarization realization of $SL\left(2, \mathbb{C} \right)$ as a Drinfel'd double.
A suitable action for the system  is the following:
\begin{equation}
\tilde{S}_0= \int_{\mathbb{R}} dt \, \tilde{L} _0= - \frac{1}{4} \int_{\mathbb{R}} {\mathcal Tr} [\tilde{\varphi}^*\left(\tilde{g}^{-1} d \tilde{g}\right) \wedge * \tilde{\varphi}^*\left(\tilde{g}^{-1} d\tilde{g}\right)] = - \frac{1}{4}\int_{\mathbb{R}} dt \,  {\mathcal Tr} [(\tilde{g}^{-1}{ \dot {\tilde{g}}})(\tilde{g}^{-1}{ \dot {\tilde{g}}})] 
\end{equation}
with $\tilde{\varphi}:t \in \mathbb{R} \rightarrow \tilde{g} \in SB(2,\mathbb{C})$, the group-valued target space coordinates, so that 
\begin{equation}
\tilde{g}^{-1} d \tilde{g}= i \tilde{\alpha}_k \tilde{e}^k 
\end{equation}
is the Maurer-Cartan left invariant one-form on the group manifold, with $\tilde{\alpha}_k$ the left-invariant basic one-forms,   $*$ is again the Hodge star operator on the source space $\mathbb{R}$ satisfying $* dt = 1$. The symbol ${\mathcal Tr} $ is used here to represent a suitable scalar product in the Lie algebra $\mathfrak{sb}(2,\mathbb{C})$. In  this case the group is not semi-simple, so there is no scalar product which is both non-degenerate and invariant. Therefore,  one has two possible different choices: the scalar product defined by the real and/or imaginary part of the trace, given by (\ref{sp1}) and (\ref{sp2}) which is $SU(2)$ and $SB(2,\mathbb{C})$ invariant but degenerate, or one could use the scalar product induced by the Riemannian metric $\mathcal{H}$, which, on the algebra $\mathfrak{sb}(2,\mathbb{C})$ takes the form $\left( \left(\tilde{e}^i, \tilde{e}^j \right) \right)=\delta^{ij}+{\epsilon^i}_{l3}\delta^{lk} {\epsilon^j}_{k3}$, positive definite and non-degenerate. However, this is $SU(2)$ invariant but only invariant under left $SB(2,\mathbb{C})$ action.
Indeed, by observing that the generators $\tilde{e}^i$ are not Hermitian, (\ref{riemannianmetric}) can be verified to be equivalent to:
\begin{equation}
((u,v)) \equiv 2 \mbox{Re}\left[ \mbox{Tr} \left(u^{\dagger} v \right)\right],
\end{equation}
so that $((\tilde{g}^{-1}{ \dot {\tilde{g}}}, \tilde{g}^{-1} \dot {\tilde{g}}))= 2\mbox{Re}\left\{ \mbox{Tr} [(\tilde{g}^{-1}{ \dot {\tilde{g}}})^{\dagger} \tilde{g}^{-1} \dot {\tilde{g}}]\right\}$ which is not invariant under right $SB(2,\mathbb{C})$ action, since  $\tilde{g}^{-1} \neq \tilde{g}^{\dagger}$. We use the latter scalar product: ${\mathcal Tr}(ab) \equiv ((a,b ))$, therefore the Lagrangian is left/right invariant under $SU(2)$ but only left $SB(2,\mathbb{C})$ invariant, while the original rotator model was invariant under left/right actions of both groups. Again the group manifold can be embedded in the $\mathbb{R}^4$ ambient space and parametrized so that $\tilde{g} \in SB(2, \mathbb{C})$ can be written as $\tilde{g}=2\left(u_0 \tilde{e}^0+i u_i \tilde{e}^i \right)$ with $\tilde{e}^0=\mathbb{I}/2$ and $u_0^2-u_3^2=1$. The latter condition follows from the $\mbox{det}\left(\tilde{g} \right)=1$ condition. This is easily understood from the explicit form of the generators, as written in (\ref{sbgen}):
\begin{eqnarray}
\tilde{e}^1= \left( \begin{array}{cc}
  0 & i \\
  0 & 0
 \end{array}  \right) ; \quad \tilde{e}^2= \left( \begin{array}{cc}
  0 & 1 \\
  0 & 0
 \end{array} \right); \quad \tilde{e}^3=\frac{i}{2} \left( \begin{array}{cc}
  1 & 0 \\
  0 & -1
  \end{array} \right)
\end{eqnarray} 
In order to be consistent we have then:
\begin{equation}
u_i=\frac{1}{4}\left( \left(i \tilde{g},\tilde{e}^i \right) \right), \, \, i=1,2, \quad \, \, u_3=\frac{1}{2}\left( \left(i \tilde{g},\tilde{e}^3 \right) \right), \quad \, \, u_0=\frac{1}{2}\left( \left( \tilde{g},\tilde{e}^0 \right) \right).
\end{equation}
Most of these calculations work in the same way as for the IRR model, with the appropriate differences in the parametrizaton of $\tilde{g} \in SB\left(2, \mathbb{C} \right)$ and in the scalar product (this time invariant only under left $SB\left(2, \mathbb{C} \right)$ action) which defines the metric $h^{ij} \equiv \left(\delta^{ij}+{\epsilon^i}_{l3}{\epsilon^j}_{k3}\delta^{lk} \right)$, hence we will not go through details in this section.
Since $\tilde{g}^{-1}=2\left(u_0 \tilde{e}^0-i u_i \tilde{e}^i \right)$ we have
\begin{equation}
\tilde{g}^{-1}\dot{\tilde{g}}=2i\left(u_0 \dot{u}_i-u_i \dot{u}_0+{f_i}^{jk}u_j \dot{u}_k \right)\tilde{e}^i=2i \dot{\tilde{Q}}_i \tilde{e}^i,
\end{equation}
where the ${f^{ij}}_{k}=\epsilon^{ijl}\epsilon_{l3k}$ are the structure constants of $\mathfrak{sb}(2,\mathbb{C})$, so that the Lagrangian can be written as
\begin{equation}
\tilde{L}_0=h^{ij}\dot{\tilde{Q}}_i \dot{\tilde{Q}}_j,
\end{equation}
having defined
\begin{equation}
\dot{\tilde{Q}}_i \equiv u_0 \dot{u}_i-u_i \dot{u}_0+{f_i}^{jk}u_j \dot{u}_k
\end{equation}
as left generalized velocities
and 
\begin{equation}\label{hmetric}
h^{ij}=\delta^{ij}+ \epsilon^{im3}\delta_{mn}\epsilon^{jn3} \,\, . 
\end{equation}
Following the same approach as with the IRR, the equations of motion of the system can be found to be
\begin{equation}
\mathcal{L}_{\Gamma} \dot{\tilde{Q}}_j h^{ji}-\dot{\tilde{Q}}_l \dot{\tilde{Q}}_m {f_k}^{il}h^{mk}=0.
\end{equation}
We can then consider $(\tilde{Q}_i, \dot{\tilde{Q}}_i)$ as tangent bundle coordinates, with $\tilde{Q}_i$ implicitly defined, similarly to the rigid rotator case.

The carrier space of the Hamiltonian dynamics is instead $T^*SB(2,\mathbb{C})$, with coordinates $(\tilde{Q}_i, \tilde{I}^i)$, with $\tilde{I}^i$ the conjugate left momenta defined as usual as
\begin{equation}
\label{dualmomentagenvel}
\tilde{I}^i=\frac{\partial \tilde{L}_0}{\partial \dot{\tilde{Q}}_i}=h^{ij}\dot{\tilde{Q}}_j.
\end{equation}
To perform the Legendre transform from $TSB(2,\mathbb{C})$ to $T^*SB(2,\mathbb{C})$ we have to invert (\ref{dualmomentagenvel}), which results in
\begin{equation}
\dot{\tilde{Q}}_i=\tilde{I}^j\left(\delta_{ji}-\frac{1}{2}\epsilon_{jp3}\epsilon_{iq3}\delta^{pq} \right),
\end{equation}
leading to the Hamiltonian
\begin{equation}
\tilde{H}_0=\frac{1}{2}\left(h^{-1} \right)_{ij}\tilde{I}^i \tilde{I}^j,
\end{equation}
being 
\begin{equation}
\left(h^{-1} \right)_{ij}=\left(\delta_{ij}-\frac{1}{2}{\epsilon_i}{p3}{\epsilon_j}^{q3} \delta_{pq} \right)
\end{equation}
 the inverse of the metric $h^{ij}$ of  (\ref{hmetric}). Similarly to what we have done  for the rigid rotator, we can introduce the linear combination $\tilde{I}=-i\tilde{I}^i {\tilde{e}}^*_i$ over the dual basis ${\tilde{e}}^*_i$, such that $\langle {\tilde{e}}^*_j | \tilde{e}^i \rangle=\delta^i_j$. Following the same steps as in the case of the rigid rotator we can find the symplectic form  from the first-order action functional, and it reads
\begin{equation}
\tilde{\omega}=d\tilde{\theta}=d \tilde{I}^i \wedge \tilde{\alpha}_i+\tilde{I}^i {f_i}^{jk}\tilde{\alpha}_j \wedge \tilde{\alpha}_k.
\end{equation}
By inverting $\tilde{\omega}$ we find the Poisson algebra
\begin{eqnarray}
\label{bracketdual}
 \{u_i, u_j \} & = & 0 \\ 
\{\tilde{I}^i, \tilde{I}^j\}  &= & {f^{ij}}_k \tilde{I}^k \\ 
\{u_i, \tilde{I}^j\} & = & \delta_i^j-{f_i}^{jk}u_k \iff \{\tilde{g} ,\tilde{I}^j\}=2i\tilde{g}\,\tilde{e}^j,
\end{eqnarray}
from which the Hamilton equations of motion can be obtained as follows:
\begin{eqnarray}
 \dot{\tilde{I}}^i &= &\{\tilde{I}^i ,\tilde{H}_0\}={f_k}^{ij}\tilde{I}^k \tilde{I}^l (h^{-1})_{jl}, \\ 
\tilde{g}^{-1}\dot{g} &= &2i \tilde{e}^i (h^{-1})_{ij} \tilde{I}^j.
\end{eqnarray}
The fact that $\tilde{I}^j$ are not conserved is expected and it expresses the non-invariance of the model under right $SB(2,\mathbb{C})$ action. One can easily check that right momenta, obtained from right-invariant vector fields which generate left action, would result to be constants of motion.

Analogously to the IRR case, we can remark that the fibers of $TSB(2,\mathbb{C})$ can be identified with $\mathfrak{sb}(2,\mathbb{C}) \simeq \mathbb{R}^3$ (as a vector space), as well as the fibers of $T^*SB(2,\mathbb{C})$, identified with the dual algebra $\mathfrak{sb}(2,\mathbb{C})^*$,  which is also isomorphic, as vector space, to $\mathbb{R}^3$, but the elements are now components of one-forms. The carrier space of the Hamiltonian dynamics for the dual model $T^*SB(2,\mathbb{C})$ is represented, as a group, by the semi-direct product of $SB(2,\mathbb{C})$ and the Abelian group $\mathbb{R}^3$, i.e. $T^* SB(2,\mathbb{C}) \simeq SB(2,\mathbb{C}) \ltimes \mathbb{R}^3$, and the Lie algebra is a semi-direct  sum represented by
\begin{eqnarray}
 \left[B_i,B_j \right] & = & i {f_{ij}}^k B_k \\ 
\left[S_i,S_j \right] & = & 0 \\ 
\left[B_i,S_j \right] & = & i {f_{ij}}^k S_k,
\end{eqnarray}
being $B_i$ the generators of the $SB(2,\mathbb{C})$ algebra and $S_i$ the generators of $\mathbb{R}^3$. Again, as before for the IRR, the non-trivial Poisson brackets (\ref{bracketdual}) can be understood in terms of the coadjoint action of $SB(2,\mathbb{C})$ on its dual algebra.

\section{The Doubled Action}
\label{secdoubled}

The two models we described can be obtained from the same parent action defined on the whole $SL(2, \mathbb{C})$ group by introducing the natural doubling of coordinates of $SL(2,\mathbb{C})$ as a Drinfel'd double. In this sense they appear as dual.

In this regard, we can introduce the following doubled notation for the $\mathfrak{sl}(2,\mathbb{C})$ generators:
\begin{equation}
e_I=\left( \begin{array}{c}
  e_i  \\
  \tilde{e}^i
   \end{array} \right) , \quad \, \, e_i \in \mathfrak{su}(2), \quad \tilde{e}^i \in \mathfrak{sb}(2,\mathbb{C}).
\end{equation}
In this notation, we can rewrite the scalar product of $\mathfrak{sl}(2,\mathbb{C})$ in (\ref{sp1}) as
\begin{equation}
\langle e_I, e_J \rangle=\eta_{IJ}= \left( \begin{array}{cc}
  0 & \delta_i^j \\
  \delta^i_j & 0
 \end{array} \right).
\end{equation}
This is the $O(3,3)$ invariant metric reproducing the fundamental structure in Doubled Geometry, i.e. the usual $O(d,d)$ invariant metric.

The Riemannian product defined in (\ref{riemannianmetric}) can be written instead as
\begin{equation}
\left( \left( e_I , e_J\right) \right)=\mathcal{H}_{IJ}= \left( \begin{array}{cc}
  \delta_{ij} & {\epsilon_{3i}}^j \\
  -{\epsilon^i}_{j3} & \delta^{ij}+{\epsilon^i}_{l3}{\epsilon^j}_{k3}\delta^{lk}
 \end{array} \right),
\end{equation}
which satisfies the relation $\mathcal{H}^{T}\eta \mathcal{H}=\eta$, indicating that $\mathcal{H}$ is a pseudo-orthogonal $O(3,3)$ matrix. In this way we can clearly see how the metric $\eta$ and the metric $\mathcal{H}$ naturally emerge out. In fact,  $\eta$ corresponds to the $O(d,d)$ invariant metric while $\mathcal{H}$ to the so-called generalized metric, with $\delta_{ij}$ playing the role of the graviton field $G_{i j}$ and $\epsilon_{ij3}$ playing the role of the Kalb-Ramond field $B_{ij}$. %as can be seen from the explicit form of the DFT generalized metric in (?).

\subsection{Lagrangian description}
In order to introduce the Lagrangian formalism of the doubled model on $TSL(2,\mathbb{C})$ we describe it again as a $0+1$-dimensional group-valued field theory, with dynamical variables  $\phi: t \in \mathbb{R} \rightarrow \gamma(t) \in SL(2,\mathbb{C})$. In particular, using the doubled coordinates we introduced before, the left-invariant Maurer-Cartan one-form on the group manifold is $\gamma^{-1}d\gamma \in \mathfrak{sl}(2,\mathbb{C}) \otimes \Omega^1(SL(2,\mathbb{C}))$ can be pulled back to $\mathbb{R}$ yielding
\begin{equation}
\phi^*\left(\gamma^{-1}d \gamma\right)=\gamma^{-1}\dot{\gamma} \,dt \equiv \dot{\textbf{Q}}^I e_I dt,
\end{equation}
being $\dot{\textbf{Q}}^I$ the left generalized velocities, which we can decompose as $\dot{\textbf{Q}}^I \equiv \left(A^i ,B_i\right)$, resulting in
\begin{equation}
\gamma^{-1}\dot{\gamma} \,dt =\left(A^i e_i+B_i \tilde{e}^i \right)dt.
\end{equation}
Both generalized velocities components are coordinates of the tangent bundle of $SL(2,\mathbb{C})$ but $\left(A^i ,B_i\right)$ could also alternatively be interpreted in terms of Generalized Geometry as fiber coordinates of the generalized bundle $T \oplus T^*$ with base space $SU(2)$.

The components of the generalized velocity can be obtained by using the scalar product (\ref{sp1}):
\begin{equation}
A^i=2\mbox{Im}\left[\mbox{Tr}\left(\gamma^{-1}\dot{\gamma}\tilde{e}^i \right)\right]; \quad B_i=2\mbox{Im}\left[\mbox{Tr}\left(\gamma^{-1}\dot{\gamma}e_i \right)\right].
\end{equation}
The doubled action on $SL(2,\mathbb{C})$ can be introduced at this point using both the scalar products, as follows:
\begin{equation}
S=\frac{1}{2}\int_{\mathbb{R}}\left[k_1\langle \phi^*\left(\gamma^{-1}d\gamma\right) , *\phi^*\left(\gamma^{-1}d\gamma \right) \rangle+k_2 \left(\left( \phi^*\left(\gamma^{-1}d\gamma\right) , *\phi^*\left(\gamma^{-1}d\gamma \right)\right) \right) \right],  \label{actiondoubled}
\end{equation}
where $k_1$ and $k_2$ are two real parameters. In terms of generalized velocities, since 
\begin{equation}
\langle \phi^*\left(\gamma^{-1}d\gamma\right) , *\phi^*\left(\gamma^{-1}d\gamma \right) \rangle=dt\, \dot{\textbf{Q}}^I \dot{\textbf{Q}}^J \langle e_I,e_J \rangle=dt\, \dot{\textbf{Q}}^I \dot{\textbf{Q}}^J \eta_{IJ}
\end{equation}
and
\begin{equation}
\left( \left( \phi^*\left(\gamma^{-1}d\gamma\right) , *\phi^*\left(\gamma^{-1}d\gamma \right) \right) \right)=dt \,\dot{\textbf{Q}}^I \dot{\textbf{Q}}^J \left( \left(e_I, e_J \right) \right)=dt \,\dot{\textbf{Q}}^I \dot{\textbf{Q}}^J \mathcal{H}_{IJ},
\end{equation}
we can write the action (up to an overall constant) explicitly in terms of the splitting of $\mathfrak{sl(2,\mathbb{C})}$ as a Drinfel'd double $\mathfrak{su}(2) \bowtie \mathfrak{sb}(2,\mathbb{C})$:
\begin{equation}
S=\frac{1}{2}\int_{\mathbb{R}}dt \, E_{IJ}\dot{\textbf{Q}}^I \dot{\textbf{Q}}^J,
\end{equation}
with $E_{IJ}=k\, \eta_{IJ}+\mathcal{H}_{IJ}$, and  we defined $k=\frac{k_1}{k_2}$. We can observe that the matrix $E_{IJ}$ is non-singular provided $k \neq 1$, which is a condition we assume from now on. Explicitly, in terms of fiber coordinates of $TSL(2,\mathbb{C})$ the Lagrangian gets the form:
\begin{equation}
L=\frac{1}{2}\left[\delta_{ij}A^i A^j+\left(k\delta_i^j+{\epsilon_i}^{j3} \right)A^i B_j+\left(k\delta_j^i-{\epsilon^i}_{j3} \right)B_i A^j+h^{ij}B_i B_j \right].
\end{equation}
The Lagrangian one-form is
\begin{equation}
\theta_L=E_{IJ} \, \dot{\textbf{Q}}^I \alpha^J,
\end{equation}
so that the equations of motion in the intrinsic formulation can be written as
\begin{equation}
\mathcal{L}_{\Gamma}\dot{\textbf{Q}}^I E_{IJ}-\dot{\textbf{Q}}^P\dot{\textbf{Q}}^Q {C_{IP}}^K E_{QK}=0,
\end{equation}
being ${C_{IP}}^K$ the structure constants of the $\mathfrak{sl}(2,\mathbb{C})$ Lie algebra.

\subsection{Hamiltonian description}

As usual, we can define the left generalized momenta in the doubled description as
\begin{equation}
\textbf{I}_I=\frac{\partial L}{\partial \dot{\textbf{Q}}^I}=E_{IJ}\dot{\textbf{Q}}^J,
\end{equation}
so that the Hamiltonian then reads as:
\begin{equation}
H=\left[\textbf{I}_I \dot{\textbf{Q}}^I-L \right]_{|\dot{\textbf{Q}}^I=(E^{-1})^{IJ}\textbf{I}_J}=\frac{1}{2}(E^{-1})^{IJ}\textbf{I}_I \textbf{I}_J
\end{equation}
with
\begin{equation}
(E^{-1})^{IJ}=\frac{1}{1-k^2} \left( \begin{array}{cc}
  \delta^{ij}+\epsilon^{il3}\delta_{lk} \epsilon^{jk3} & -\delta_{pj}\left(\epsilon^{ip3}+k\delta^{ip}\right) \\
 \delta^{pj}\left(\epsilon_{ip3}-k\delta_{ip}\right) & \delta_{ij}
 \end{array} \right).
\end{equation}
Moreover, we can write the components of $\textbf{I}_I$ explicitly in terms of the $(A^i, B_i)$ components:
\begin{equation}
\textbf{I}_{I} \equiv (I_i, \tilde{I}^i)=\left(\delta_{ij}A^j+\left(k\delta_i^j+{\epsilon_i}^{j3} \right)B_j, \left(k \delta^i_j-{\epsilon^i}_{j3} \right)A^j+\left(\delta^{ij}+\delta^{lk}{\epsilon^i}_{l3}{\epsilon^j}_{k3} \right)B_j \right),
\end{equation}
so that in terms of the components $(I_i, \tilde{I}^i)$ the Hamiltonian can be written as follows:
\begin{equation}
H=\frac{1}{2\left(1-k^2 \right)}\left[\left(\delta^{ij}+\epsilon^{il3}\delta_{lk}\epsilon^{jk3} \right)I_i I_j+\delta_{ij}\tilde{I}^i \tilde{I}^j-2\left(\epsilon^{ip3}+k\delta^{ip} \right)\delta_{pj}I_i\tilde{I}^j \right].
\end{equation}
We can consider the linear combination $\textbf{I}=-\frac{i}{2} \, \textbf{I}_I e^{I^*}=-\frac{i}{2}\left(I_i e^{i^*}+\tilde{I}_i \tilde{e}^{*}_i \right)$, such that, using also $\gamma^{-1}d \gamma=2i \, \textbf{$\alpha$}^K e_K=i\left(\alpha^k e_k+\beta_k \tilde{e}^k \right)$ we obtain the symplectic form $\textbf{$\omega$}$ on $T^* SL(2, \mathbb{C})$:
\begin{equation}
\textbf{$\omega$}=d \textbf{$\theta$}=dI_i \wedge \textbf{$\alpha$}^i+\textbf{I}_I \, {C^{I}}_{JK}\, \textbf{$\alpha$}^J \wedge \textbf{$\alpha$}^K,
\end{equation}
from which the Poisson brackets for the generalized momenta can be obtained:
\begin{eqnarray}
\label{poissgenermom}
 \{I_i, I_j \} & = & {\epsilon_{ij}}^k I_k \\ 
\{\tilde{I}^i , \tilde{I}^j\} & = & {f^{ij}}_k \tilde{I}^k \\ 
 \{I_i, \tilde{I}^j \} & = & {\epsilon_{il}}^j \tilde{I}^l-I_l {f_i}^{lj}, \quad \{\tilde{I}^i, I_j \}=-{\epsilon^i}_{jl}\tilde{I}^l+I_l {f^{li}}_j ,
\end{eqnarray}
while those between momenta and configuration space variables are unchanged with respect to $T^* SU(2)$ and $T^* SB(2,\mathbb{C})$. We can write these brackets in compact (and doubled) form as
\begin{equation}
\{\textbf{I}_I, \textbf{I}_J \}={C_{IJ}}^K \textbf{I}_K.
\end{equation}
Finally, the Hamilton equations can be derived as follows:
\begin{equation}
\dot{\textbf{I}}_I=\{\textbf{I}_I, H \}=(E^{-1})^{JK}\{\textbf{I}_I ,\textbf{I}_J \} \textbf{I}_K=(E^{-1})^{JK} {C_{IJ}}^L \, \textbf{I}_L \textbf{I}_K.
\end{equation}

\subsection{Relation with Generalized Geometry and Poisson-Lie symmetries}

Since we can consider the isomorphism $TSL(2,\mathbb{C}) \simeq SL(2,\mathbb{C}) \times \mathfrak{sl}(2,\mathbb{C})$ with the fiber
\begin{equation}
\mathfrak{sl}(2,\mathbb{C}) \simeq \mathfrak{su}(2) \oplus \mathfrak{sb}(2,\mathbb{C}) \simeq TSU(2) \oplus T^*SU(2),
\end{equation}
one can rewrite the Poisson algebra (\ref{poissgenermom}) as
\begin{equation}
\label{cbrack}
\{I+\tilde{I}, J+\tilde{J} \}=\{I, J \}-\{J, \tilde{I} \}+\{I, \tilde{J}\}+\{\tilde{I}, \tilde{J} \},
\end{equation}
having defined $I=i I_i e^{i^*}$, $J=iJ_i e^{i^*}$ as one-forms and $\tilde{I}=\tilde{I}^i \tilde{e}^*_i$, $\tilde{J}=\tilde{J}^i \tilde{e}^*_i$ as vector fields.
Poisson brackets (\ref{poissgenermom}) are given by the KSK brackets on the coadjoint orbits of $SL(2,\mathbb{C})$, but in particular, they are induced by the bialgebra structure of $SL(2,\mathbb{C})$ and according to (\ref{cbrack}) they can be identified with the \textit{C-brackets}  \cite{Deser2015,Deser2018} of Generalized geometry \footnote{C-brackets are mixed brackets between vector fields and forms.
They generalize Courant and Dorfmann brackets}
for the generalized bundle $T \oplus T^*$, being $\{e^{i^*}\}$ and $\{\tilde{e}^*_i \}$ bases over $T^*$ and over $T$ respectively. Namely, the doubled momenta $(I_i, \tilde{I}^i)$ identify the fiber coordinates of the generalized bundle $T \oplus T^*$ of $SU(2)$. 

Furthermore, defining Hamiltonian vector fields in terms of Poisson brackets as usual as
\begin{equation}
X_f = \{\cdot, f \}
\end{equation}
and defining in particular $X_i=\{\cdot, I_i \}$, $\tilde{X}^i=\{\cdot, \tilde{I}^i \}$, one can find, because of the non-trivial Poisson bracket (\ref{poissgenermom}), and by using Jacobi identity:
\begin{eqnarray}
 [X_i, X_j ] & = & \{\{\cdot,I_j\},I_i\}-\{\{\cdot,I_i\},I_j\}=\{\cdot ,\{I_i,I_j\}\}={\epsilon_{ij}}^k X_k, \\ 
\left[ \tilde{X}^i, \tilde{X}^j \right] & = & \{\{\cdot,\tilde{I}^j\},\tilde{I}^i\}-\{\{\cdot,\tilde{I}^i\},\tilde{I}^j\}=\{\cdot ,\{\tilde{I}^i,\tilde{I}^j\}\}={f^{ij}}_k \tilde{X}^k,\\ 
\left[ X_i, \tilde{X}^j \right] & = & \{\{\cdot,\tilde{I}^j\},I_i\}-\{\{\cdot,I_i\},\tilde{I}^j\}=\{\cdot ,\{I_i,\tilde{I}^j\}\}=-{f_i}^{jk} X_k-\tilde{X}^k {\epsilon_{ki}}^j,
\end{eqnarray}
or, in a unified fashion:
\begin{equation}
[X+\tilde{X}, Y+\tilde{Y}]=[X,Y]+[\tilde{X}, \tilde{Y}]+\mathcal{L}_X \tilde{Y}-\mathcal{L}_Y \tilde{X}.
\end{equation}
This shows, remarkably, that the $C$-brackets can be obtained as derived brackets \textcolor{red} {\cite{Deser2018a}}  from the canonical Poisson brackets of the dynamics.
\vspace{1.1mm}

It is important at this point to summarize and discuss in what sense the two submodels possess Poisson-Lie symmetries. We have seen in Sect. \ref{secdrinfpoissonlie} what Poisson-Lie T-duality means and how it is related to the concept of Drinfel'd double. Namely, we have seen that under appropriate conditions the sigma models defined on groups that are dual partners in a Manin triple polarization are indeed dual, in the sense that they describe the same physics even if there is no such manifest symmetry in neither of the two dual actions. Indeed, the two models can be seen to be connected by a canonical transformation on their phase-space variables and classically their dynamics is indistinguishable. However, a parent model can be formulated on the Drinfel'd double group and at this stage Poisson-Lie duality becomes a manifest symmetry and the two submodels can be obtained by gauging conditions. Furthermore, there are two symmetric ways to perform the decomposition: $\gamma= \tilde{g}g$ or $\gamma=g \tilde{g}$. 
In our simple case, we started from the action of an isotropic rigid rotator on the group manifold $SU(2)$, and having realized that $SL(2,\mathbb{C})$ can be seen as a Drinfel'd double and in particular $\mathfrak{sl}(2,\mathbb{C})= \mathfrak{su}(2) \bowtie \mathfrak{sb}(2,\mathbb{C})$ is a Manin triple, we built the dual model on $SB(2,\mathbb{C})$, they can then be obtained from the generalized action on $SL(2,\mathbb{C})$ under appropriate gauging. In this sense we can see we have the ingredients under which Poisson-Lie duality relies. It is already enough in principle to state that the model is a Poisson-Lie model. 
\\
However, note that in this case the model is too simple to have Poisson-Lie symmetry, indeed, there does not exist a canonical transformation, that is due to the fact  the model under analysis is not a genuine field theory. In fact, this can be found instead in the principal chiral model case \cite{Marotta2019}.

\subsection{Recovering the dual models}
\label{secrecover}

The standard dynamics of the isotropic rigid rotator and its dual model can be recovered from the doubled Lagrangian we have introduced. In order to get back one of the two models one has to impose constraints, as is customary in DFT. In particular, one has to gauge either $SU(2)$ or $SB(2,\mathbb{C})$ and integrate out.

For definiteness, we specify and fix a local Iwasawa decomposition for the elements of $SL(2,\mathbb{C})$ as $\gamma=\tilde{g}g$, with $\tilde{g} \in SB(2,\mathbb{C})$ and $g \in SU(2)$. From the action in (\ref{actiondoubled}) and the properties we have remarked on the two scalar products defined on $SL(2,\mathbb{C})$, it can be seen that the Lagrangian is manifestly globally invariant under both left and right $SU(2)$ actions but only under left $SB(2,\mathbb{C})$ action. Therefore, in order to recover the $TSU(2)$ rotator description this left $SB(2,\mathbb{C})$ invariance has to be promoted to a gauge symmetry and then gauged appropriately. The left $SB(2,\mathbb{C})$ action is given by
\begin{equation}
SB(2,\mathbb{C})_L : \gamma \rightarrow \tilde{h}\gamma=\tilde{h}\tilde{g}g, \quad \forall \, \tilde{h} \in SB(2,\mathbb{C}).
\end{equation}
Promoting this global symmetry to a gauge one, we modify the Maurer-Cartan one-form defining the covariant exterior derivative $D_{\tilde{C}}=d+\tilde{C}$, where $\tilde{C}$ is the gauge connection one-form $\tilde{C}=\tilde{C}_i(t) \tilde{e}^i$, so that
\begin{equation}
\phi^*\left(\gamma^{-1}d \gamma \right) \rightarrow \phi^*\left(\gamma^{-1}D_{\tilde{C}}\gamma \right)=\left(\gamma^{-1}\dot{\gamma}+\gamma^{-1} \tilde{C}\gamma \right) dt.
\end{equation}
We can make explicit the doubled notation by performing the following splitting:
\begin{equation}
\label{splitrecov}
\gamma^{-1}\dot{\gamma}+\gamma^{-1} \tilde{C}\gamma=U_i \, \tilde{e}^i+W^i e_i,
\end{equation}
whose components can be obtained from
\begin{equation}
U_i=2 \mbox{Im}\left\{\mbox{Tr}\left[\left(\gamma^{-1}\dot{\gamma}+\tilde{C}_j \gamma^{-1}\tilde{e}^j \gamma \right)e_i \right] \right\}
\end{equation}
and
\begin{equation}
W^i=2 \mbox{Im}\left\{\mbox{Tr}\left[\left(\gamma^{-1}\dot{\gamma}+\tilde{C}_j \gamma^{-1}\tilde{e}^j \gamma \right)\tilde{e}^i \right] \right\}.
\end{equation}
These can be computed explicitly by means the coadjoint action of $g$, $\tilde{g}$ on $e_j$, $\tilde{e}^j$ represented by the Lie brackets (\ref{liesl2c}) (for the $e_i$), (\ref{liesb}) and (\ref{liesusb}). However, this is not necessary for our purposes and details can be found in \cite{Marotta2018}. In terms of these new degrees of freedom we can write down the doubled Lagrangian with the gauge connection added, so the gauged Lagrangian reads
\begin{eqnarray}
L_{\tilde{C}} =\frac{1}{2}\left[\delta_{ij}W^i W^j+2\left(k\delta_i^j+{\epsilon_i}^{j3} \right)W^i U_j+h^{ij}U_i U_j \right],
\end{eqnarray}
since
\begin{eqnarray}
L_{\tilde{C}} {} & =\frac{1}{2} \left[ k_1\langle \phi^* \left(\gamma^{-1}D_{\tilde{C}}\gamma \right) , *\phi^*\left( \gamma^{-1}D_{\tilde{C}}\gamma\right) \rangle+k_2 \left(\left( \phi^* \left(\gamma^{-1}D_{\tilde{C}}\gamma \right) , *\phi^*\left( \gamma^{-1}D_{\tilde{C}}\gamma\right)  \right) \right) \right] \\ & =\frac{1}{2} E_{IJ}\dot{\hat{\textbf{Q}}}^I\dot{\hat{\textbf{Q}}}^J,
\end{eqnarray}
with $\dot{\hat{\textbf{Q}}}^I=(W^i,U_i)$. Then, performing the transformation
\begin{equation}
\hat{W}^i=W^i+\left(k\delta^{is}-{\epsilon_3}^{is} \right)U_s,
\end{equation}
we have
\begin{equation}
L_{\tilde{C}}=\frac{1}{2}\left[\delta_{ij}\hat{W}^i \hat{W}^j+\left(1-k^2 \right)\delta^{ij}U_i U_j \right].
\end{equation}
We will use this form for writing the Euclidean partition function of the system
\begin{equation}
\mathcal{Z}=\int \mathcal{D}_g \mathcal{D}_{\tilde{g}}\mathcal{D}_{\tilde{C}} \, e^{-S_{\tilde{C}}} 
\end{equation}
and integrate over the gauge connection. In particular, we can trade the integration over $\tilde{C}_i$ with an integration over $U_i$:
\begin{equation}
\mathcal{Z}=\int \mathcal{D}_g \mathcal{D}_{\tilde{g}} \, \mbox{det}\left(\frac{\delta \tilde{C}_i}{\delta U_j} \right) e^{-\frac{1}{2}\int_{\mathbb{R}}dt \, \delta_{ij}\hat{W}^i \hat{W}^j} \int \mathcal{D}_U e^{-\frac{1}{2}\int_{\mathbb{R}} dt \,\left(1-k^2 \right)\delta^{ij}U_i U_j }.
\end{equation}
It is easy to see that the Jacobian determinant of the $\tilde{C} \rightarrow U$ transformation is constant since the matrices involved in the gauge transformations are constant, hence it only results in a regularization factor. Using the fact that the Gaussian integral over $U$ is also a constant, the partition function can be finally written in the form
\begin{equation}
\mathcal{Z} \propto \int \mathcal{D}_g \mathcal{D}_{\tilde{g}}  e^{-\frac{1}{2}\int_{\mathbb{R}}dt \, \delta_{ij}\hat{W}^i \hat{W}^j}.
\end{equation}
In order to compare with the IRR action (\ref{S0action}) we can make a step further. It is possible to introduce the endomorphism $E$ of $\mathfrak{sl}(2,\mathbb{C})=\mathfrak{su}(2) \oplus \mathfrak{sb}(2, \mathbb{C})$ which preserves the Drinfel'd splitting, defined by the constant matrix
\begin{equation}
E= \left( \begin{array}{cc}
  \delta^i_j & T^{ij} \\
  -(T^{-1})_{ij} & \delta_i^j
 \end{array} \right)
\end{equation}
such that we can make the following splitting-preserving change of variables on $\mathfrak{sl}(2,\mathbb{C})$ as a Drinfel'd double:
\begin{equation}
E \left( \begin{array}{c}
  W^j \\
  U_j
 \end{array} \right) = \left( \begin{array}{c}
  \hat{W}^j \\
  \hat{U}_j
 \end{array} \right).
\end{equation}
In this way, we can write the Maurer-Cartan left-invariant one-forms as
\begin{equation}
\Phi^*\left({g'}^{-1}dg'\right)=\hat{W}^i e_i dt, \quad \Phi^*\left(\tilde{g}'^{-1} d\tilde{g}'\right)=\hat{U}_i \tilde{e}^i dt.
\end{equation}
The endomorphism $E$ induces an exponential map $\mbox{exp}(E):SL(2,\mathbb{C}) \rightarrow SL(2,\mathbb{C})$ such that $\gamma=\tilde{g}g$ is mapped into $\gamma'=\tilde{g}' g'$, so that the integration measure can be transformed into $\mathcal{D}_{g'} \mathcal{D}_{\tilde{g}'}$, hence up to a constant factor (the determinant of $\mbox{exp}(E)$) the partition function can be written as
\begin{equation}
\mathcal{Z}\propto\int\mathcal{D}_{\tilde{g}'}\int \mathcal{D}_{g'} e^{-\frac{1}{2}\int_{\mathbb{R}}\mbox{Tr}\left[\Phi^*\left(g'^{-1}d g' \right) \wedge * \Phi^*\left(g'^{-1}d g' \right)\right]}.
\end{equation}
Clearly the integration over $\tilde{g}'$ gives another constant, while the other integral is the partition function of the IRR model.

The dual model with carrier space $TSB(2,\mathbb{C})$ can be recovered following exactly the same procedure but gauging this time the global right $SU(2)$ action invariance. The main difference with respect to the previous case is that the gauge connection one-form is now $\mathfrak{su}(2)$-valued, and under the integral it is suitable to be traded for the integration over the dual analogue of $W$ in (\ref{splitrecov}). This case has been carried out in detail in \cite{Marotta2018}. 

\section{Conclusions}

After reviewing some of the fundamental aspects of Drinfel'd double and Poisson-Lie T-duality, we have considered the dynamics of the  three-dimensional isotropic rigid rotator on the group manifold of $SU(2)$ and its dual model on the group $SB(2,\mathbb{C})$, first introduced in \cite{Marotta2018}, in the spirit of outlining their connection with Poisson-Lie sigma models. 
%on $SL(2,\mathbb{C})$,  which is  a non-Abelian deformation of the natural phase space of the  rotator dynamics,  $T^*SU(2) \simeq SU(2) \ltimes \mathbb{R}^3$ \cite{MARMO1995}. From this,  a   new model has been introduced in \cite{Marotta2018} on the dual group $SB(2,\mathbb{C})$, which we have  in the sense that it is the dual partner of the Drinfel'd double description of $SL(2,\mathbb{C})=SU(2) \bowtie SB(2,\mathbb{C})$. 
We have analyzed the two models from the Poisson-Lie duality point of view, building a doubled generalized model with $TSL(2,\mathbb{C})$ as carrier space. This was done with the purpose of exploring more deeply the relations between Poisson-Lie symmetries, Double Geometry and Generalized Geometry in a particularly simple system so that the framework could be more easily and explicitly understood. In fact, we were able to recover all the mentioned structures in such a simple system; for example, we found a Poisson realization of the C-brackets for the generalized bundle $T \oplus T^*$ over $SU(2)$ from the Poisson algebra of the generalized model. This shows that C-brackets can be obtained as derived brackets, in analogy with the ideas presented in \cite{Deser2015,Deser2018},  with the remarkable property that in this case they are derived from the Poisson brackets of the dynamics.

The two dual models exhibit many  characteristics of dual Poisson-Lie sigma models and from the generalized action both can be recovered by gauging one of its symmetries, as it is customary in the framework of Double Field Theory. However, the dynamics described by the two models is not equivalent. 

DFT has the purpose of making T-duality manifest in the target space low-energy effective theory. 
%DFT can be considered indeed a manifestly T-dual invariant reformulation of Supergravity. 
Many attempts have been done in the direction of generalizing the DFT description to Poisson-Lie T-duality, which is a generalization of the standard Abelian and non-Abelian concepts of T-dualities, see e.g. \cite{Hassler2017}.
However, a manifest implementation of T-duality already at the level of world-sheet string action has not been yet clearly recovered, which would be of course even more interesting. The isotropic rigid rotator is a toy model, representing  a sigma model in $0+1$ dimensions. Its true field theory generalization, which is the Principal Chiral  Model, is presented in  \cite{Marotta2019}; the inclusion into the latter of a Wess-Zumino term is currently under investigation \cite{WZW}.  
Besides being interesting per se, these models may contribute to a better understanding   of the  string world-sheet formulation, with the possibility of writing  a manifestly $O(d,d)$ T-duality invariant doubled 
%string world-sheet 
action  in a more rigorous way
%, keeping in mind that the configuration space is no longer a Lie group, but a differentiable manifold.  
and then performing the low energy limit where all the results obtained in Double Field Theory should be recovered. 

\vspace{.2cm}

{\bf Acknowledgments} F.P. would like to thank the organizers of the Workshop on {\em Dualities and Generalized Geometries} for the invitation to deliver this talk.

\end{document}